\documentclass[fleqn,usenatbib]{mnras}

\usepackage[T1]{fontenc}

\DeclareRobustCommand{\VAN}[3]{#2}
\let\VANthebibliography\thebibliography
\def\thebibliography{\DeclareRobustCommand{\VAN}[3]{##3}\VANthebibliography}

\usepackage{graphicx}	
\usepackage{amssymb}	
\usepackage{amsmath}	
\usepackage{color}
\usepackage[english]{babel}
\usepackage{url}
\usepackage{ulem}
\usepackage{multirow}
\usepackage{hyperref}

\usepackage{newtxtext,newtxmath}

\title[Mixing Sinc kernels in SPH]{Mixing Sinc kernels to improve interpolations in smoothed particle hydrodynamics without pairing instability}

\author[R. Cabezón \& D. García-Senz]{
Rubén M. Cabezón,$^{1}$\thanks{E-mail: ruben.cabezon@unibas.ch}
and Domingo García-Senz,$^{2,3}$
\\
$^{1}$Center for Scientific Computing (sciCORE). Universit\"at Basel, Klingelbergstrasse 61, 4056 Basel, Switzerland\\
$^{2}$Departament de F\'{i}sica, Universitat Polit\`ecnica de Catalunya, Avinguda Eduard Maristany 16, E-08019 Barcelona (Spain)\\
$^{3}$Institut d'Estudis Espacials de Catalunya, Gran Capit\` a 2-4, E-08034
Barcelona, Spain
}

\date{Accepted XXX. Received YYY; in original form ZZZ}

\pubyear{2023}

\begin{document}
\label{firstpage}
\pagerange{\pageref{firstpage}--\pageref{lastpage}}
\maketitle

\begin{abstract}
The smoothed particle hydrodynamic technique is strongly based on the proper choice of interpolation functions. This statement is particularly relevant for the study of subsonic fluxes and turbulence, where inherent small errors in the averaging procedures introduce excessive damping on the smallest scales. To mitigate these errors we can increase both the number of interpolating points and the order of the interpolating kernel function. However, this approach leads to a higher computational burden across all fluid regions. Ideally, the development of a single kernel function capable of effectively accommodating varying numbers of interpolating points in different fluid regions, providing good resolution and minimal errors would be highly desirable. In this work, we revisit and extend the main properties of a family of interpolators called {\it Sinc kernels} and compare them with the widely used family of Wendland kernels. We show that a linear combination of low- and high-order Sinc kernels generates good-quality interpolators, which are resistant to pairing instability while maintaining good sampling properties in a wide range of neighbor interpolating points, $60\le n_b\le 400$. We show that a particular case of this linear mix of Sincs produces a well-balanced and robust kernel that improves previous results in the Gresho-Chan vortex experiment even when the number of neighbors is not large, while yielding a good convergence rate. Although such a mixing technique is ideally suited for Sinc kernels owing to their excellent flexibility, it can be easily applied to other interpolating families such as the B-splines and Wendland kernels.
\end{abstract}

\begin{keywords}
methods: numerical -- hydrodynamics
\end{keywords}



\section{Introduction}
Despite the enormous progress made by the smoothed particle hydrodynamics (SPH) method since it was formulated \citep{lucy77,gingold77} there is still room for improvement in different areas. Among them, convergence issues are of utmost importance. Especially if we want to grasp fine details in fluid-dynamics studies, such as numerically reproducing three-dimensional turbulence \citep{price18}. In SPH, the best approach is to use a Lagrangian-compatible formulation \cite{spr02} plus a large number of particles and neighbors \citep{zhu15}, the two latter being limited by the available computational resources. However, details matter, such as how gradient estimation is implemented (either traditional \citep{mon92,pri12}, least squares \citep{dilts99}, or integral \citep{garciasenz2012} approaches) and the quality of the interpolating kernel \citep{fulk96,ros15}.

Traditionally, the SPH technique has made use of the B-splines family of polynomials, $M_n$, to carry out the interpolations that are at the foundation of the method. This is a well-justified choice because these kernels have compact support, are fast to compute, and have some flexibility in the choice of the interpolating polynomial, going from the widely used cubic spline $M_4$ to higher-order $M_n$~($n=5,6,7 ...$) polynomials. The $M_n$ family has a support interval $R_n$ that is not the same among all the kernels, being larger $R_n$ for higher-degree polynomials. As a consequence, the number of neighbors allocated within $R_n$ has to increase as $n$ increases, to keep a good sampling of the fluid when interpolating. Otherwise, the scarcity of neighbor particles close to the center of the kernel makes the density estimation too dependent on the particle self-contribution, then inducing what is commonly known as {\it bias} \citep{deh12,ros15,price18}.
Furthermore, the $M_n$ family has a nonpositive Fourier transform $\mathcal {F}_d$, where $d$ is the number of dimensions. As a consequence, they can fall prey to the so-called pairing (or clustering) instability when the number of neighbors is high. The existence of the pairing instability was first noted by \cite{sch81} who realized that above a critical number of neighbor particles, many compact-supported kernels tend to produce an artificial particle clumping, which not only degrades the effective resolution but produces nonphysical gradients of pressure.

A different family of polynomials $C^n$, based on Wendland functions \cite{wendland1995} and more resistant to the pairing instability, was fully discussed by \cite{deh12} in the context of the SPH technique. Unlike the $M_n$ family, these $C^n$ (or Wendland) kernels have $\mathcal{F}(C^n)_d (k)> 0, \forall k$, where $k$ is the wavenumber. Nevertheless, its practical execution demands taking a high number of neighbors to elude bias. However, increasing the number of neighbors raises the computational effort considerably, not only in terms of the number of operations per particle but also in terms of communications overhead in massively distributed calculations.

Finally, another kind of interpolator is the Sinc kernel family, $S_n$ \citep{cabezon2008, garciasenz2014}, directly linked to the Dirac-$\delta$ function. These are a very flexible uniparametric family of interpolators steered by the kernel exponent $n$, with similar properties as the B-splines family. Although $\mathcal {F}_d (S_n)$ oscillates around zero, an appropriate choice of the exponent $n$ shifts $\mathcal{F}_d (k) < 0$ to high wavenumbers, thus suppressing the pairing instability in practice.

In this work we discuss more deeply the properties of the $S_n$ family of kernels proposed in \citep{cabezon2008}, with emphasis on the connections between the value of the governing exponent $n$ of the kernel, its Fourier transform $\mathcal {F}_d (S_n)$, and the quality of the sampling. We show that combining linearly two $S_n$ kernels with different exponents gives rise to interpolators that are even more resistant to particle clustering while retaining good sampling properties. A correct sampling also helps to control the so-called E0 errors which in SPH appear whenever the integrals defining the average of physical magnitudes and its gradients are approached by finite summations \citep{mon05}. These E0 errors result in SPH not achieving complete second-order accuracy. Nevertheless, they can be minimized by a correct choice of the interpolation kernel and by increasing the number of interpolating points. The suggested linear mix of low- and high-order kernels can be used either when the number of interpolating neighbor particles is small, allowing for higher spatial resolution with less computational effort, or if the number of neighbors is large, reducing the E0 errors without suffering undesirable particle clustering.

Our main objective is therefore to build a general-purpose kernel able to handle an ample range of neighbor particles without suffering from particle clustering and with good sampling properties. According to our numerical experiments, mixing two Sinc kernels can achieve this result while still rivaling the best results obtained with several members of the $C^n$ family of polynomials. But unlike the $C^n$ or $M_n$ kernel families, our default mixed Sinc kernel does not require a previous and careful choice of the kernel order as a function of the chosen number of target neighbors. For example, our study suggests that the Wendland $C^2$ behaves better than $C^4$ and $C^6$ when the number of neighbors is low or moderate, $n_b\le 150$, but it is recommended to turn it to $C^4$ and $C^6$ progressively in the interval $150\le n_b\le 400$. As we will see, this shortcoming is outpowered by the novel interpolator proposed in this work, with no additional inconveniences.

The main properties of Sinc and mixed Sinc kernels are reviewed in Sects.~\ref{sec:kernels} and \ref{sec:mixing}. Section \ref{sec:convergence} presents tests that highlight the ability of mixed Sinc kernels to accurately reproduce the density profile in a uniform system (Sect.~{\ref{sec:zeroth_test}}), to overcome the pairing-instability, (Sect.~\ref{sec:first_test}), and to handle the Gresho-Chan vortex experiment (Sect.~\ref{sec:second_test}) and the Sod shock-tube test (Sect.~\ref{sec:third_test}). We summarize the main results of our work and provide some guidelines for future work in the Conclusions Section.

\section{Properties of the Sinc kernels}
\label{sec:kernels}

The Sinc interpolation family, $S_n$, was first proposed by \cite{cabezon2008} as a practical and flexible way to switch between interpolators of different widths and sharpness with a single parameter. They are spherically symmetric functions that come directly from one of the numerical representations of a Dirac-$\delta$ function \citep{cabezon2017},

\begin{equation}
S_n=D_n \left[\frac{\sin{(\frac {\pi}{2}} u)}{\frac{\pi}{2}u}\right]^n,
\label{Sincs}
\end{equation}

\noindent where $D_n$ is a normalization constant, $u=\vert \mathbf{r}'-\mathbf{r}\vert/{h}$ is the particle-neighbor distance normalized to the local smoothing length $h$, and $n$ is the kernel index. The $S_n$ family has compact support in $u$, defined in the interval $[0:2]$~with radius\footnote{The radius $R$ is defined as the distance to the kernel center at which $S_n=0$ and $\partial(S_n)/\partial u=0$.} $R=2$, and its main features have been described in \citep{cabezon2008, garciasenz2014, cabezon2017, ros15}. One of the aims of the original work was to suppress the pairing instability by simply raising the value of the kernel exponent. Increasing $n$ makes the kernel sharper, thus shifting $\partial^2 S_n/\partial{u^2} =0$ to low $u$ values. It turns out that raising the index of the kernel also pushes $\mathcal F_d (S_n) >0 $ to high wave numbers, which hinders particle clustering.

\begin{table}
        \centering
        \caption{Main features of Sinc ($S_n$) and Wendland ($C^n$) interpolators. $S^{\alpha}_{n,m}$ stands for the mixed Sinc interpolators described in the text. Columns are: (from left to right) exponents/order of the interpolating kernel (from top to bottom, pure Sinc, mixed Sinc, and Wendland kernels), the standard deviation $\sigma$ (Eq.~\ref{std}), the effective smoothing length $\ell$ according to \citet{deh12}, the interval of compact support radius $R=2$  normalized to $\ell$, the full-width half maximum of the interpolator, and the location of the first zero ($k_c$) of the Fourier transform of the kernel.}
        \begin{tabular}{cccccc}
                \hline
                $S_n$ & $\sigma$ & $\ell= 2\sigma$ & $R/\ell$& FWHM & $k_c/2\pi$\\
                \hline
                \hline
                3 & 0.540 & 1.080 & 1.852&1.468&0.840\\
                4 & 0.487 & 0.973 & 2.055&1.276&1.029\\
                5 & 0.446 & 0.892 & 2.242&1.148&1.224\\
                6 & 0.414 & 0.828 & 2.415&1.048&1.425\\
                7 &0.388 & 0.776 & 2.579&0.972&1.638\\
                8 &0.366 & 0.732 & 2.732&0.912&1.860\\
                9 &0.348 & 0.696 & 2.877&0.860&2.304\\
                \hline
                Mixed & $\sigma$ & $\ell= 2\sigma$ & $R/\ell$& FWHM & $k_c/2\pi$\\
                \hline
                \hline
                $S^{0.9}_{4,9}$ &0.475 & 0.950 & 2.105&1.152&1.935\\
                $S^{0.9}_{5,9}$ &0.437 & 0.874 & 2.288&1.080&2.082\\
                $S^{0.4}_{5,6}$ &0.427 & 0.854 & 2.342&1.080&1.680\\
                \hline
                $C^n$ & $\sigma$ & $\ell=2\sigma$ & $R/\ell$& FWHM& $k_c/2\pi$\\
                \hline
                \hline
                2 & 0.516 & 1.032 & 1.938&1.256& -\\
                4 & 0.354 & 0.708 & 2.829&1.124& -\\
                6 & 0.261 & 0.522 & 3.830&1.012& -\\
                \hline
                \hline
        \end{tabular}
        \label{table1}
\end{table}

Table~\ref{table1} shows several features of the $S_n$ and $C^n$ families with compact support $[0:2]$, such as the standard deviation,
\begin{figure*}
\centering
\includegraphics[width=\textwidth]{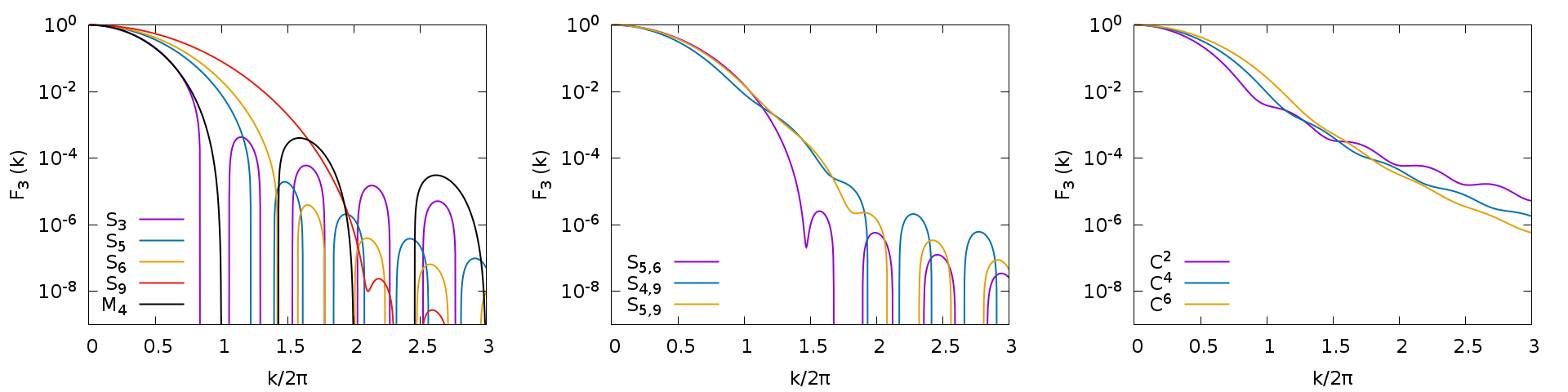}
\caption{Fourier transforms of several kernels: Sinc ($S_n$, left panel), mixed $S_n$ (middle panel), and Wendland ($C^n$, right panel). }
\label{fig:FT_1}
\end{figure*}

\begin{figure*}
\centering
\includegraphics[width=\textwidth]{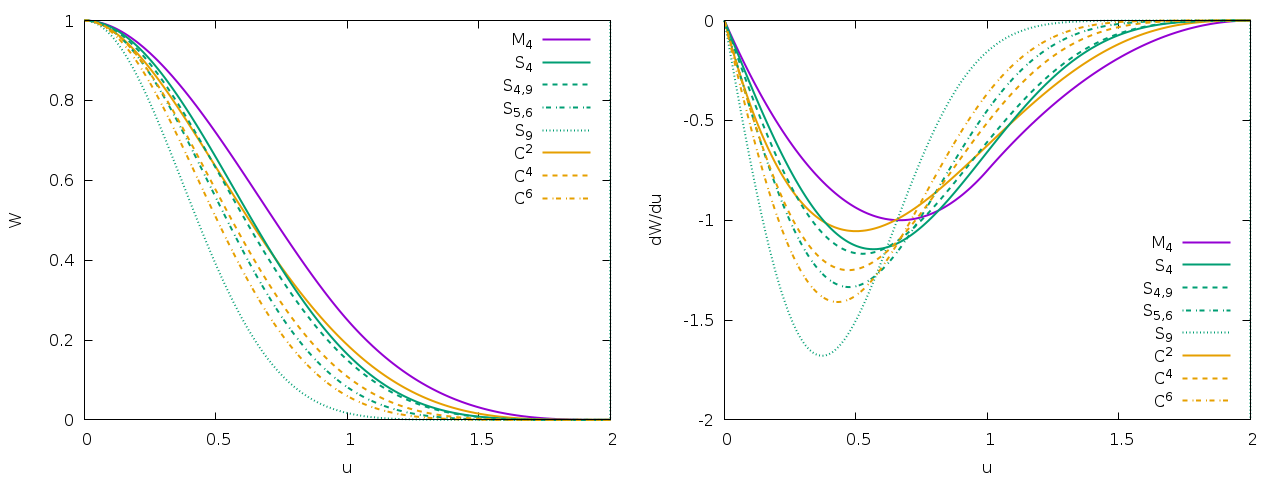}
\caption{Radial profiles of several kernels discussed in the text (left) and their first derivative (right). We did not apply any normalization to ease the comparison. We assign the same color to kernels that belong to the same family, Sinc or Wendland (the B-spline $M_4$ is also shown), while the type of line depicts different kernels within the same family.}
\label{fig:profiles}
\end{figure*}
\begin{equation}
\sigma=1/3\int_{\infty}{\mathbf x}^2 W({\mathbf x},h) d^3{\mathbf x},
\label{std}
\end{equation}

\noindent and the distance $\ell=2\sigma\simeq 1$ (roughly at a half of $R$) that is often taken as the effective smoothing length \citep{deh12}. The last column shows the full width at half maximum (FWHM) of the kernels, which we found is a useful parameter to make comparisons among them. For example, the properties of the $S_3$ interpolator shown in Table~\ref{table1} are close to those of the cubic spline, $M_4$, kernel. It has the same support interval $[0:2]$ and similar $\sigma (M_4)=0.548$ and FWHM$(M_4)=1.448$ values. A comparison between the $M_n$, $C^n$, and $S_n$ profiles in \citep{cabezon2017} suggests the following rough analogy: $\{S_3, S_4, S_5\}\equiv\{M_4, M_5, M_6\}$ and $\{S_4, S_5, S_6\}\equiv \{C^2, C^4, C^6\}$, respectively. The similarity between these Sinc and Wendlands interpolators on a purely morphological basis is also supported by the similarity among their FWHM values, as shown in the fifth column of Table~\ref{table1}. Nevertheless, a second parameter related to the Fourier Transform (FT) of the interpolating function is also necessary to better characterize kernels. This is because, despite having similar shapes, their FT can be very different and, more specifically, its behavior at a large wavenumber.
\begin{figure*}
\centering
\includegraphics[width=\textwidth]{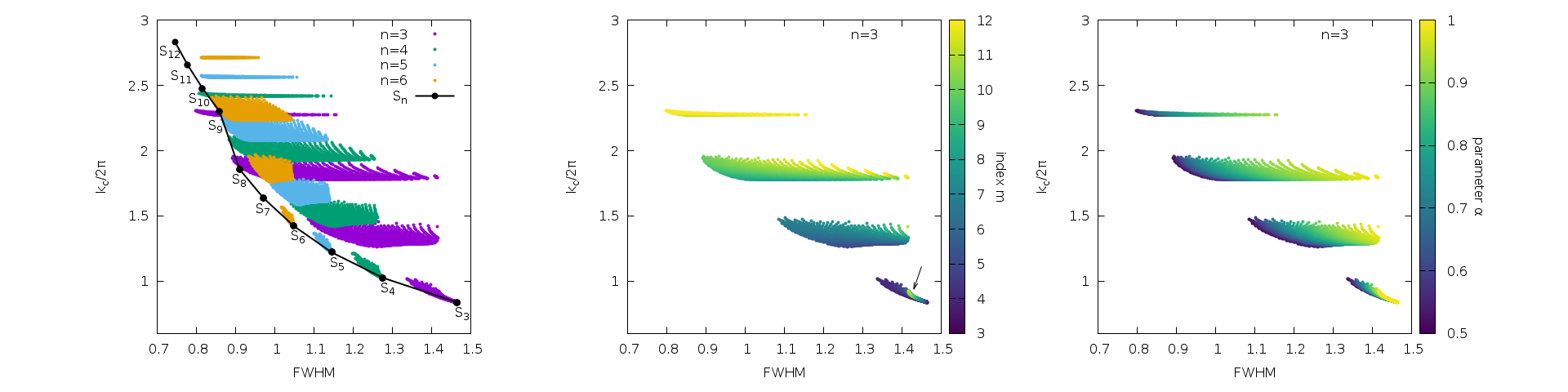}
\caption{Depiction of the influence of the kernel mixing (left), the Sinc index $m$ (center), and the $\alpha$ parameter (right) on $k_c$ (i.e. pairing resistance) and FWHM (i.e. bias resistance). The solid line shows the $k_c$-FWHM evolution for pure unmixed Sinc kernels with integer values $3\le n\le 12$. The panels in the center and right show the effects only for the case of a low-order kernel $n=3$ in the mixing. The arrow is clarified in the text.}
\label{fig:indexes}
\end{figure*}
\begin{equation}
    \mathcal{F}_3[w(r)](k)=4\pi
    k^{-1}\int_{0}^{\infty} \sin(k r) w(r) r dr\,,
    \label{fourier}
\end{equation}

\noindent where $w(r)$ is a spherically symmetric kernel and $k$ is the wavenumber in the three-dimensional space. We show the Fourier transforms of several $S_n$ kernels, $\mathcal{F}(S_n)(k)$ in the range $3\le n\le 9$ calculated with Eq.~(\ref{fourier}) in the left panel of Fig.~\ref{fig:FT_1}. As noted in previous works \citep{garciasenz2014, cabezon2017}, $\mathcal{F}(S_n)(k)$ remains positive at long wavenumber when $n$ is high but begins to oscillate around zero after a critical wavenumber $k_c$, which we show for several kernels in the last column of Table~\ref{table1}. Therefore, the FT of the $S_n$ family is non-positive definite. We also show in the same panel of Fig.~\ref{fig:FT_1} the FT of the $M_4$ polynomial that oscillates strongly after $k>k_c$. Nevertheless, high-order kernels $S_n$ and $M_n$ manage to shift $k_c$ to higher wavenumbers so that they behave, in practice, closer to kernels with positive-definite Fourier transform. For instance, according to Fig.~\ref{fig:FT_1} (left panel), the $S_9$ interpolator remains positive in a range of wavelengths that nearly triples that of $S_3$. On another note, the $C^n$ family is constructed to have $\mathcal{F}(S_n)(k)>0$ for all wavenumbers (right panel of Fig.~\ref{fig:FT_1}). Because avoiding artificial particle clustering (i.e., pairing instability) is bounded to a positive FT of the kernel \citep{deh12}, the $C^n$ polynomials have become very popular in the field, especially when using a large number of neighbor particles. The central panel of Fig.~\ref{fig:FT_1} shows the effect of mixing Sinc kernels on the shape of the FT, pushing $k_c$ towards larger wavenumbers. This will be discussed in more detail in Sect.~\ref{sec:mixing}.

The other side of the coin is that high-order kernels of any kind suffer, in general, from bias when the number of neighbors is not large enough. If the number of neighbors is low and the kernel is sharp, the sampling within the kernel range may become inadequate owing to the loss of statistical weight in the central region of the kernel when compared to the particle self-contribution. In these cases, the best option is to use a not-too-sharp kernel such as, for example, $M_5$, $S_5$, or $C^2$. On the contrary, higher-order interpolators are necessary if the number of neighbors is large, which reduces the E0 errors, but drastically increases the computational effort and also the chances of suffering from particle clustering if the kernel is not pairing resistant. Additionally, problems may arise if the number of neighbors fluctuates too much during the simulation runtime, as errors can grow larger in regions where the neighbor count decreases.

A plausible solution to the above dilemma is to consider a mix of Sinc kernels that simultaneously retain the good sampling properties of low-order kernels (i.e. with low exponents $n$) and the known abilities of high-order kernels (a large $n$ exponent) to elude the pairing instability. We propose to build interpolators $S^{\alpha}_{n,m}$ with $n\le m$, such as

\begin{equation}
S^{\alpha}_{n,m} = \alpha S_n + (1-\alpha) S_m
\label{Slinear_mix}
\end{equation}

\noindent where $0\le \alpha\le 1$ is a mixing parameter. For example, taking $\alpha=0.9$ and mixing Sincs $S_4$ and $S_9$ leads to $S^{0.9}_{4,9}$. The resulting interpolators are spherically symmetric, fulfill the normalization constraint and, therefore, share all the conservation properties of standard SPH kernels. There are a huge number of combinations, but we are interested in those fulfilling the following ad-hoc requirements:

a) The Fourier transform of $S^{\alpha}_{n,m}$ has its first zero closer to (or even beyond) that of $S_m$ (see the central panel in Fig.~\ref{fig:FT_1}). This guarantees that the ensuing kernel has a similar pairing resistance as the high-order kernel part of the mix.

b) The FWMH of $S^{\alpha}_{n,m}$ should be similar to that of Sincs with low or moderate exponents. This guarantees a good sampling even when the number of neighbors is low.

Figure~\ref{fig:profiles} depicts the radial profiles of several kernels discussed in this work (left) and their radial derivative (right). It is worth noting that, despite having remarkably similar profiles and FWHM values, kernels $S_4$ and $S_{4,9}^{0.9}$ yield considerably different results in the tests below. This highlights the relevance of the details of the interpolating kernel.

We note that, unlike other kernel families, the pairs $(n,m)$ do not necessarily have to be integer numbers, which provides vast flexibility. The computational overload introduced by the linear mixing procedure is totally negligible if the kernel value and its derivatives are previously stored in a table from which the necessary interpolations are performed during the hydrodynamic calculation.

\begin{figure*}
\centering
\includegraphics[width=\textwidth]{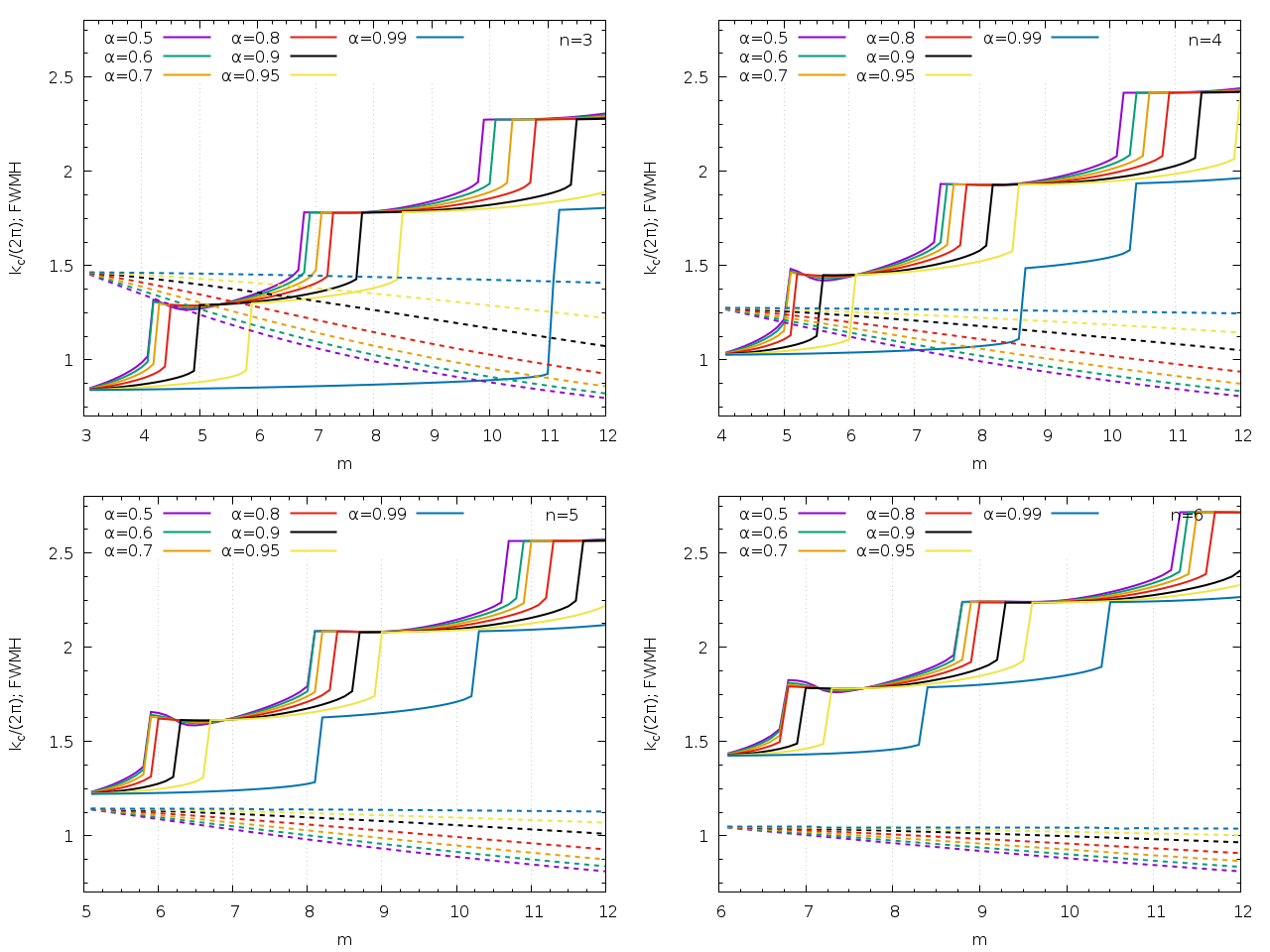}
\caption{FWHM and zeroes of the FT of mixed Sinc kernels as a function of the high-order kernel index $m$ and the mixing parameter $\alpha$. The panels show the results for different integer low-order kernel exponents $n$. The exponent $m$ is varied in steps of 0.1 and $m>n$. The vertical axes show the critical wavenumber bounded to the first zero of the FT (solid lines) and the FWHM (dashed lines).}
\label{fig:zeroes}
\end{figure*}

\section{Understanding the mixing}
\label{sec:mixing}
As mentioned in the previous section, our main metrics to have a first assessment of the quality of a kernel are the values of $k_c$ and FWHM. Ideally, a kernel with high values on both would be a good candidate for a pairing-resistant and bias-insensitive kernel that is accurate within a wide range of numbers of neighbors. With this in mind, it is important to understand how the mixing of Sinc kernels proposed in Sect.~\ref{sec:kernels} affects $k_c$ and FWHM. There are three aspects to explore: the general effect of the mixing, the value of the exponent of the high-order kernel ($m$), and the value of the mixing parameter ($\alpha$).

Figure~\ref{fig:indexes} provides an insight into the effect of each parameter. In all cases, the mixing parameter $\alpha$ is varied in the range $0.50\le \alpha\le 0.99$ in steps of $0.01$. On the leftmost panel, we show the values of $k_c$ as a function of FWHM for the Sinc family $S_n$ with $3\le n \le 6$ (on the solid line). That is our starting situation. It is clear here that as we increase the index $n$ of the kernel, it becomes more pairing-resistant as its $k_c$ increases, but at the expense of reducing FWHM, as expected for more peaked kernels. Our hypothesis is that such anti-correlation is at the root of the issues discussed in Sect.~\ref{sec:kernels}. Finally, it is worth noting the notable jump in $k_c$ from $S_8$ to $S_9$. We discuss this feature in more detail below.

The effect of mixing these $S_n$ kernels with $S_m$ kernels ($n\le m \le 12$) is shown by the colored points. For the sake of clarity, we show in this plot only the cases where $n$ is an integer ($3\le n\le 6$) and $m$ is a varying real ($n+\delta \le m\le 12.00$, with $\delta=0.01$). Each color shows the metrics of all linear combinations that have the corresponding $n$ as the low-index kernel, for all values of $m$ and $\alpha$. As can be seen, the mixing pushes $k_c$ to higher values still at the cost of reducing FWHM, but considerably less for most combinations than simply increasing the exponent $n$ in the unmixed kernels. The fact that the solid line behaves as a limit to the left and below almost all colored points signals that the proposed mixing should produce better interpolating kernels.

The second and third panels of Fig.~\ref{fig:indexes} focus on the individual effect of the index $m$ and the parameter $\alpha$ in the mixing. Here, we show only the results assuming $n=3$, but the behavior is the same for all $n$. It is clear that increasing $m$ is the reason for a higher $k_c$ while increasing $\alpha$ produces larger FWHM. The exception to this is the bright line on the bottom right of these plots (marked with an arrow in the central panel), which corresponds to $\alpha=0.99$. In this case, the weight of the low-order kernel is too high and brings the metrics back, close to the original value of the solid line on the left panel, this being the limit achieved at $\alpha=1$.
In this way, we confirm our initial assumptions that a higher-order kernel with high index $m$ is needed to increase pairing resistance, but not too high to reduce the FWHM too much. In addition, a high $\alpha$ is needed to compensate for the loss of FWHM when mixing with a high $m$, but not too high to lose the influence of the high-order kernel on $k_c$.

\begin{figure}
\centering
\includegraphics[width=\columnwidth]{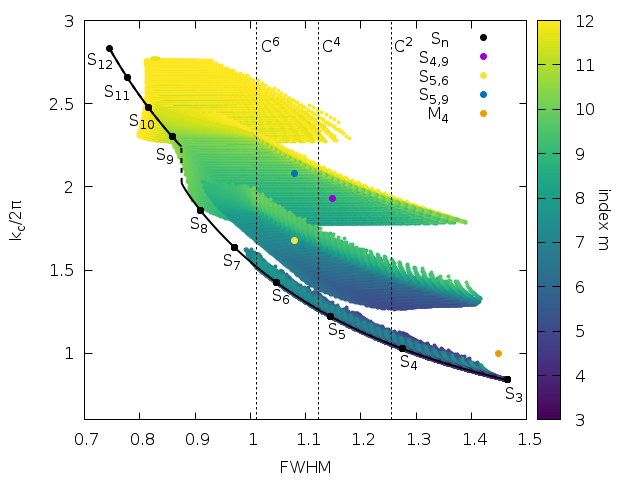}
\caption{$k_c$ in function of FWHM for the pure Sinc family (black solid line) and all combinations $S_{n,m}^{\alpha}$, for $0.5\le\alpha\le0.98$, $3\le n\le 6$, and $n+\delta\le m\le 12$, with $\delta=0.01$. We used the same step $\delta$ for all three parameters. The vertical dashed lines show the FWHM of the Wendland kernels.}
\label{fig:all}
\end{figure}

Figure~\ref{fig:zeroes} shows the evolution of $k_c$ and FWHM for several combinations $(n,m)$ of Sincs and different $\alpha$. This provides certain information that guided our choice of values for $n$, $m$, and $\alpha$ over the vast parameter space:
\begin{itemize}
    \item Increasing $n$ leads to larger $k_c$ (shifting solid lines up in the panels) but lower FWHM (shifting dashed lines down) due to their anticorrelation. As expected, when the kernels become more peaked, their FWHM decreases.
    \item If $\alpha$ is too close to 1 (solid blue lines), $k_c$ is significantly degraded, coming close to the values corresponding to the unmixed Sinc with exponent $n$.
    \item If $\alpha$ is too low (dashed purple lines), the FWHM is degraded as it becomes dominated by the higher-order kernel of the mix.
    \item The FWHM values are less sensitive to $\alpha$ for increasing $n$, as seen from the lower spread of the dashed lines. If $n$ is already high, mixing it with even higher $m$ will not have a large effect on their FWHM independently of the value of $\alpha$, for the range studied.
    \item For a given $n$ and $\alpha$, low values of $m$ have the largest FWHM but the lowest $k_c$, which are similar to the values corresponding to the unmixed $S_n$.
    \item For a given $n$ and $\alpha$, large values of $m$ considerably improve $k_c$ and degrade FWHM, as expected for extremely peaked kernels.
    \item For a given $n$ and $\alpha$, large jumps in $k_c$ occur at specific $m$ values.
\end{itemize}

The strong modulation in $k_c$ that produces the wave-like pattern driven by $m$, as seen in Fig.~\ref{fig:zeroes} is due to two mechanisms: local phase compensation in consecutive $(n,m)$ pairs and overpowering of the highest exponent kernel when $m >> n$. The first is well represented by the combination of Sincs $S^{0.4}_{5,6}$ shown in the central panel of Fig.~\ref{fig:FT_1}. In this case, the negative gap of $S_5$ starting at $k/2\pi\simeq 1.2$, shown in the left panel, is compensated by the positive value of $S_6$, while the negative value of $S_6$ starting at $k/2\pi \simeq 1.4$ is balanced by the positive value of $S_5$ in that region. As shown in Table \ref{table1}, this compensation leads to a notable jump in the first zero of the Fourier transform. However, a much more efficient shifting in $k_c$ results from $m>>n$ as, for example, $S^{0.9}_{4,9}$. According to the left and middle panels in Fig.~\ref{fig:FT_1}, the new interpolator has its first zero very close to that of the Sinc with the highest exponent $m=9$. However, the FWHM of $S^{0.9}_{4,9}$ is still close to that of $S_4$ (actually, it becomes very similar to the FWHM of $S_5$) which reduces the bias. The combination of these effects produces the collapse of the first negative zone of the FT and boosts the value of $k_c$. This is also an inherent property of the Sinc kernels, which is discussed below.

If we initially focus on integer values of $n$ and $m$, Fig.~\ref{fig:zeroes} and the points discussed above will considerably limit our choices. We are mainly interested in balanced interpolators that combine good sampling properties (large FWHM) with good pairing resistance (high value of $k_c$). Hence, $n=4$ and $\alpha=0.9$ (black lines in Fig.~\ref{fig:zeroes}) are a reasonable starting point. The case with $\alpha=0.95$ (yellow lines in Fig.~\ref{fig:zeroes}) could also be another possibility, but we prioritized the slightly higher $k_c$ of the case $\alpha=0.90$. The chosen value of $m$ should be high, preferably after a jump in $k_c$, but not too high that the FWHM is degraded. This leaves $m=9$ as a preference choice.

As a result, we analyze in detail the behavior of mixed $S^{0.9}_{4,9}$, which shows remarkably good properties in a wide range of values of interpolating neighbor particles, $60\le n_b\le 400$. We also explore additional cases, such as $S^{0.4}_{5,6}$ and $S^{0.9}_{5,9}$, to test our assumptions.

\begin{figure}
\centering
\includegraphics[width=\columnwidth]{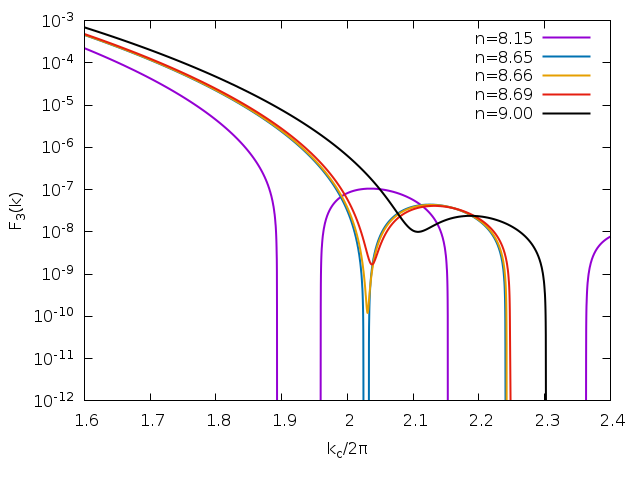}
\caption{Changes in the FT of the pure Sinc kernels as the index $n$ increases. Between $n=8.65$ and $n=8.66$ the first negative region of the FT collapses, effectively boosting $k_c$ to the next FT zero at a higher wavenumber.}
\label{fig:FT_displacement}
\end{figure}

Figure~\ref{fig:all} summarizes $k_c$ as a function of FWHM for all combinations $S_{n,m}^{\alpha}$ discussed above, with the value of $m$ color-coded. We also show the FWHM values of the Wendland kernels as vertical dashed lines. Additionally, we highlight the values corresponding to the $M_4$ kernel for comparison, as well as the discussed mixed Sinc kernels $S_{4,9}^{0.9}$, $S_{5,6}^{0.4}$, and $S_{5,9}^{0.9}$. We note that in this figure, the solid black line for the pure $S_n$ family now shows the data for all real exponents $3\leq n\leq 12$ in steps of $\delta=0.01$, instead of the data for just the integer values of index $n$, as in Fig.~\ref{fig:indexes}. In these fine-grain data, a discontinuity appears between the indexes $n=8$ and $n=9$. Specifically at $n=8.65$. Investigating further the evolution of the shape of the FT with respect to index $n$, we unveiled an interesting feature of the Sinc kernels. Figure~\ref{fig:FT_displacement} shows a detail of the FT for different indexes $n$ around their respective $k_c$. This reveals that increasing $n$ has two effects: pushing $k_c$ to higher values and {\it reducing the width} of the first negative region of the FT. Between $n=8.65$ and $n=8.66$ the first negative region of the FT collapses, which effectively causes $k_c$ to jump to considerably higher values. For this reason, we used $m=9$ in many of our mixed Sincs, as this is the first integer index value that benefits from this boost in $k_c$.

\begin{figure}
\centering
\includegraphics[width=\columnwidth]{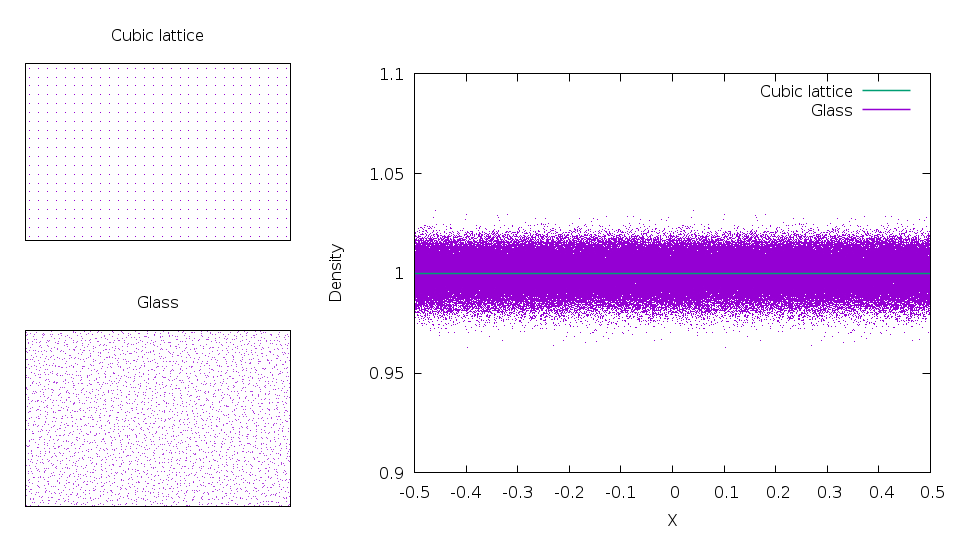}
\caption{Left column shows a comparison of the particle distribution between a cubic lattice and a glass-like configuration. The panel on the right shows the density for each configuration.}
\label{fig:lattice_comparison}
\end{figure}

\section{Tests}
\label{sec:convergence}
\begin{figure*}
\centering
\includegraphics[width=\textwidth]{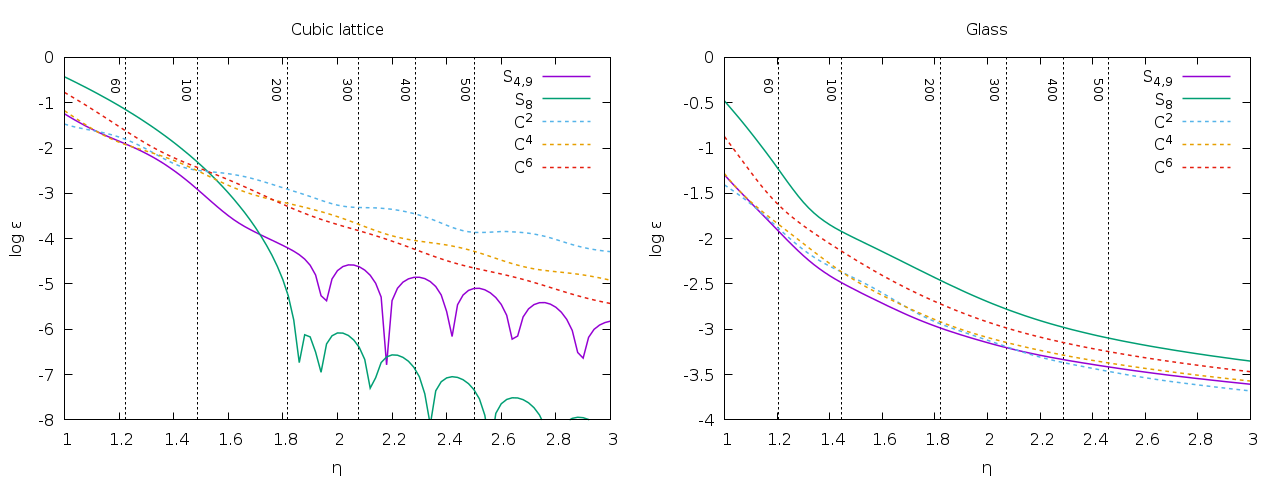}
\caption{Static box accuracy test for a cubic lattice (left) and a glass-like (right) configuration. The vertical dashed lines show the corresponding $\eta$ for several average neighbor counts.}
\label{fig:static_box}
\end{figure*}
We describe here four test cases specifically aimed at comparing the abilities of Sincs and Wendland interpolators in terms of accuracy and handling both the pairing instability and E0-like errors. These are the density accuracy in a static box \citep{ros15}, the stability of a uniform system initially in pressure equilibrium, the dynamic stability of a fluid vortex (Gresho-Chan test) \cite{gresho1990}, and the shock-tube test \citep{sod1978}. When necessary, we use an ideal gas equation of state, $P=(\gamma-1)\rho u$, with $\gamma=5/3$. The vortex test has been extensively used to analyze the convergence properties of SPH codes in the past \cite{spr10, read2012}, and the shock-tube test focuses on the kernel ability to handle fluid discontinuities. Other comparative tests on the behavior of the different families of interpolators can be found in \citep{cabezon2008,deh12,ros15,val16}. Simulations were carried out with the code SPHYNX\footnote{\url{https://astro.physik.unibas.ch/sphynx}}  \citep{cabezon2017} including the updates discussed in \citep{garciasenz2022}.

\subsection{Density accuracy in a static box}
\label{sec:zeroth_test}
In this test, we follow \cite{ros15,rosswog2023}, homogeneously distributing $100^3$ particles with identical mass in a 3D box with length $L=1$ and periodic boundary conditions. The mass of each particle is chosen so that the theoretical density of the system is $\rho_0=1$. We use two different approaches that are commonly employed in SPH simulations to achieve a homogeneous-like distribution: either distributing the particles on a cubic lattice or a relaxed glass-like configuration. The lattice approach has the advantage that the initial density of the system can be perfectly mapped by the particle distribution, albeit a lattice is not a physically consistent representation of particle distributions in most production simulations. A less artificial distribution is achieved by a glass configuration, where in order to attain no preferred directions a few percent of the density homogeneity is sacrificed (see Fig.~\ref{fig:lattice_comparison}).

Once the initial distribution is fixed, we calculate the SPH density of each particle, changing the interpolation kernel and its smoothing length. The change in the smoothing length is associated with an increasing number of neighboring particles and it is controlled by a parameter ($\eta$) that fixes the proportionality of the smoothing length with the local interparticle distance as follows:
\begin{equation}
    h_a=\eta V_a^{1/3}\,,
    \label{eta}
\end{equation}
\noindent
where $V_a=V_{box}/N$, the volume of the whole domain is $V_{box}$, and $N$ is the total number of particles.
For each given $h$, a neighbor search provides the particles needed to evaluate the SPH density $\rho_{a}$. Next, we compute the average relative error as:
\begin{equation}
    \epsilon=\frac{1}{N}\sum_{a=1}^N \frac{|\rho_a-\rho_0|}{\rho_0}
\end{equation}

Figure~\ref{fig:static_box} shows the results for a pure Sinc kernel, $S_8$, our chosen combination $S^{0.9}_{4,9}$, Wendland kernels $C^2$, $C^4$, and $C^6$. The panel on the left shows that our results are in perfect agreement with those of \cite{rosswog2023} (see their Fig.~1), which were also performed in a cubic lattice. The only new curve here is that of our mixed kernel $S^{0.9}_{4,9}$. From these results, it is clear that, in the case of a cubic lattice, Sinc kernels (either pure or mixed) outperform Wendland kernels by several orders of magnitude in the region of $n_b\gtrsim 150$. As expected for peaked kernels, $S_8$ and $C^6$ see their accuracy hindered at a low neighbor count. Only $S^{0.9}_{4,9}$ is consistently able to be among the most accurate kernels in all neighbor counts. The fact that $S^{0.9}_{4,9}$ is dominated by $S_4$ somewhat reduces its accuracy compared to $S_8$ for large $n_b$ in this setting, but it still provides considerably higher accuracy than all Wendland kernels.

The right panel of Fig.~\ref{fig:static_box} shows the results of the same test, now performed on a glass-like particle distribution \cite{arth2019}. In this case, given that the initial dispersion of the density profile is larger, the error is higher for all kernels than in the case of a cubic lattice. Nevertheless, this case is, in fact, more representative for the particle disorder that can be found on actual numerical simulations, with one caveat: this test requires fixing the smoothing length first and finding the corresponding neighbors afterward. This is in opposition to the traditional approach (used on the following tests in this work) of first fixing the number of neighbors and then adapting the smoothing length to it. However, given the homogeneity of this scenario, we consider that it is still a close representation of real calculations.

In this case, the results show that tests in a cubic lattice might be misleading, as the accuracy hierarchy seen in the cubic lattice is mostly reverted in the glass-like configuration. Also, although there are fewer differences between the kernel accuracies in the latter, $S^{0.9}_{4,9}$ is the only kernel that performs well independently of the neighbor count and the initial particle setting, either in a lattice or in a glass-like distribution.

\begin{figure*}
\centering
\includegraphics[width=0.7\textwidth]{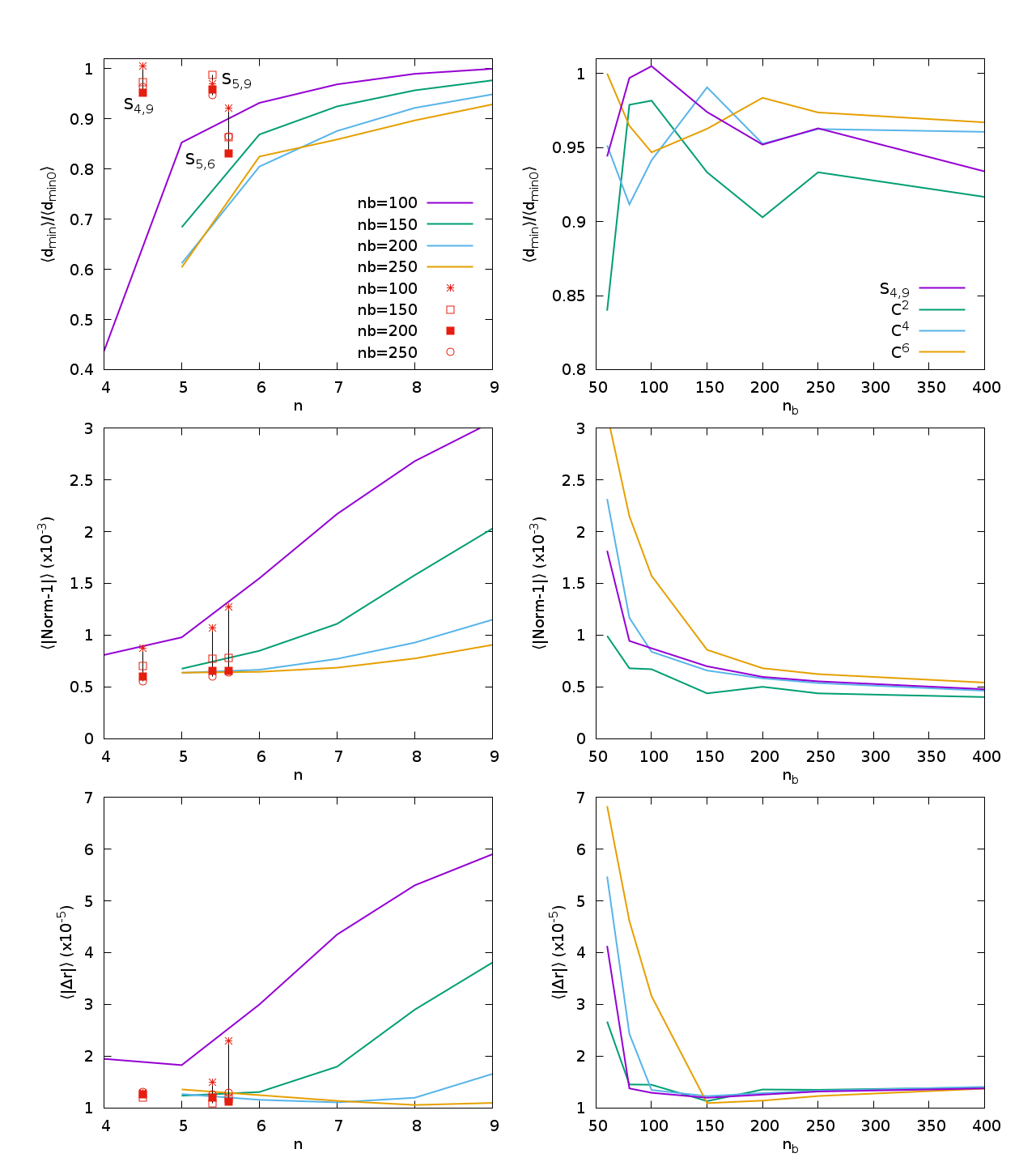}
\caption{Comparison between Sinc ($4\le n\le 9$; left column) and Wendland ($C^2, C^4, C^6$; right column) kernels in light of several indicators shown in the vertical axis of each row. The first row of panels depicts the average of the normalized minimum interparticle distance at one crossing time, $t\simeq 1$, as a function of the kernel exponent, $n$ (left), and the target number of neighbors $n_b$ (right). Red points denote the three mixed Sinc kernels, which are abbreviated as $S_{4,9}\equiv S^{0.9}_{4,9}$, $S_{5,9}\equiv S^{0.9}_{5,9}$, and $S_{5,6}\equiv S^{0.4}_{5,6}$, and they are arbitrarily located at non-integer exponents $n=4.5$, $n=5.4$, and $n=5.6$, respectively. The upper right panel shows the same indicator but for several $C^n$ polynomials and the $S_{4,9}$ mixed Sinc kernel as a function of the target number of neighbors. The second row shows the averaged error in the normalization condition, while the third row presents the degree of fulfillment of the $\left< \Delta r\right>=0$ condition. }
\label{tensile_1}
\end{figure*}

\begin{figure*}
\centering
\includegraphics[width=\textwidth]{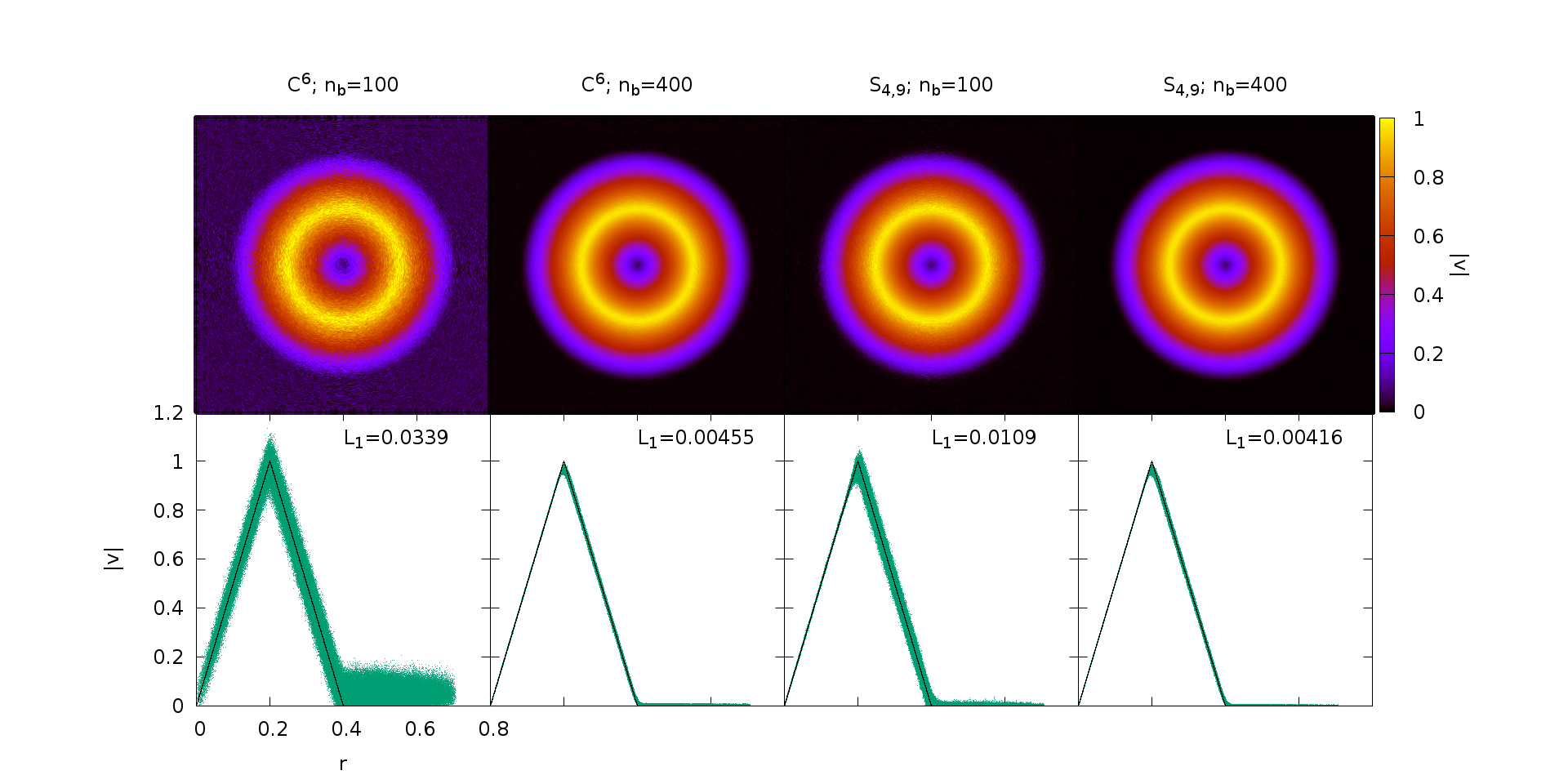}
\caption{Velocity colormap (top row) and $L_1$ errors of the azimuthal velocity (bottom row) in the Gresho-Chan vortex experiment at $t=1$, for the highest resolution simulation ($N_{1D}=400$), number of neighbors $n_b=100, 400$, and interpolators $C^6$ and $S^{0.9}_{4,9}$. The black solid line is the analytical value (Eq.~\ref{vortex_1}).}
\label{fig:Vortex_0}
\end{figure*}

\begin{figure*}
\centering
\includegraphics[width=0.7\textwidth]{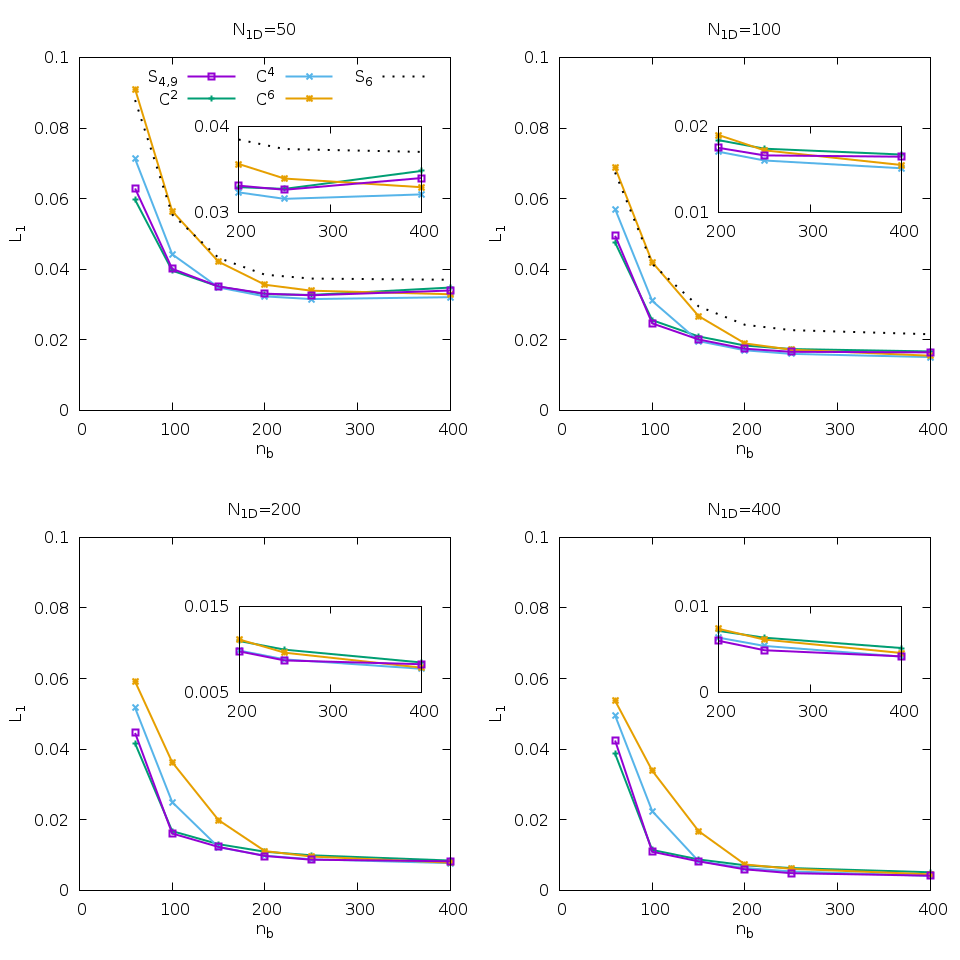}
\caption{$L_1$ errors of the azimuthal velocity field in the Gresho-Chan vortex test. The total number of particles used in each snapshot is $N_{1D}^3$ and the target neighbors set at the beginning of the simulation is $50\le n_b \le 400$. Each panel includes an inset zooming in the region of large neighbor count.}
\label{fig:Vortex_1}
\end{figure*}

\subsection{Homogeneous system in pressure equilibrium}

This test consists of a homogeneous system with $\rho_0=1$, in pressure equilibrium, $P_0=1$, in a cubic box with length $L=1$ and periodic boundary conditions. We consider $N=100^3$ particles with identical mass arranged in a glass-like configuration and follow their evolution after one crossing time $t=1$. We then analyze the results as a function of the chosen interpolator and the initial number of neighbors of each particle.

We have analyzed the quality of the results at $t=1$ in light of three estimators: a) the average of the minimum interparticle distance normalized to its initial value, $\left< d_{min}\right>/\left<d_{min0}\right>$, b) the fulfillment of the normalization condition $\sum_b V_b W_{ab}(h_a)=1 $, expressed as the averaged $\left<\sum_b V_b W_{ab}(h_a)-1\right>$ and c) the fulfillment of $ \Delta {\mathbf r} = \sum_b  (\mathbf{r}_a-\mathbf{r}_b) W_{ab}(h_a)\simeq 0$ condition expressed as the value of the average of $ \left<\vert \Delta {\mathbf r}\vert \right>$. Averages were estimated in the central region of the box with radius $L/2$. Figure~\ref{tensile_1} shows a summary of the results. The panels on the left depict the profile of these estimators as a function of the exponent $n$ of pure $S_n$ Sinc kernels. The symbols with different point types in red are for mixed kernels $S^{0.9}_{4,9}$, $S^{0.9}_{5,9}$, and $S^{0.4}_{5,6}$ from left to right, respectively. The desired behavior would be to have $\left< d_{min}\right>/\left< d_{min0}\right>\simeq 1$ (that is, no pairing), $\left<\sum_b V_b W_{ab}(h_a)-1\right>\simeq 0$, and $ \left<\vert \Delta {\mathbf r}\vert \right>\simeq 0$ (ensuring a good partition of unity and low E0 errors, respectively). Looking at the results, several conclusions can be drawn. 1) There is an inverse correlation between particle clustering and E0 errors, with low-order kernels showing lower errors but significant pairing as $n_b$ increases. 2) As expected, the mixed interpolators are always located in the desired region of the figures with both low particle clustering and small sampling errors. Among them, the kernel $S^{0.9}_{4,9}$ shows the best behavior, confirming our decision of choosing it as default to perform the numerical experiments with the Gresho-Chan vortex described in Sect.~\ref{sec:second_test} and the Sod tube in Sect.~\ref{sec:third_test}. Being a homogeneous system, the partition of the unity (middle row panels in Fig.~\ref{tensile_1}) and the $\left< \Delta r \right>=0$ condition (bottom row panels) are in general well preserved. Both conditions are better fulfilled by low-order interpolators, although above $n_b\simeq 200$ all kernels converge to similar values. However, mixed Sinc interpolators, especially $S^{0.9}_{4,9}$ and $S^{0.9}_{5,9}$ lead to the best overall results.

The right column of Fig.~\ref{tensile_1} aims at comparing the performance of $S^{0.9}_{4,9}$ with that of the $C^2$, $C^4$, and $C^6$ Wendland kernels. The results suggest that the mixed Sinc behaves better than $C^4$ and $C^6$ for \mbox{$n_b\le 100$} and $n_b\le 150$, respectively (middle and lower panels), with similar results above $n_b > 150-200$. Wendland $C^2$ gives smaller errors at $n_b < 100$, but the interparticle distance seems to oscillate more in that region (upper panel).

\label{sec:first_test}

\subsection{The Gresho-Chan vortex}
\label{sec:second_test}
The simulation of a fluid vortex in equilibrium under pressure and inertial forces, popularly known as the Gresho-Chan vortex \citep{gresho1990}, is a demanding test for any hydrodynamic code. Although the conservation of angular momentum is almost perfect in SPH codes, the results with regard to this test initially showed a low convergence rate to the analytical results \cite{spr10, deh12}. More recent works \citep{read2012,ros15, cabezon2017} have managed to increase the convergence rate as the number of particles increases. Improvements in artificial viscosity (AV) algorithms \cite{cul10}, the use of high-order kernels \citep{deh12, ros15}, and the introduction of new procedures to estimate gradients \cite{read2012, garciasenz2012, cabezon2017, hu2014} were especially relevant to successfully cope with this scenario.

We want to investigate whether the implementation of the mixed kernel $S^{0.9}_{4,9}$ in the latest version of the code SPHYNX leads to additional benefits for this test. We estimate and compare the convergence rate with the results obtained with interpolators $C^2$, $C^4$, and $C^6$.

In this test, we use the common initial conditions for the azimuthal velocity and pressure profile:

\begin{equation}
  v_\phi(r)=v_0
\begin{cases}
    \psi & \text{for $\psi\leq 1$}\\
    2-\psi & \text{for $1<\psi\leq 2$}\\
    0 & \text{for $\psi>2$,}\\
\end{cases}
\label{vortex_1}
\end{equation}

\begin{equation}
  P(r)= P_0+4v_0^2
\begin{cases}
    \psi^2/8 & \text{for $\psi\leq 1$}\\
    \left(\psi^2/8-\psi +\ln \psi +1\right) & \text{for $1<\psi \leq 2$}\\
    \left(\ln 2 -1/2\right) & \text{for $\psi>2$,}\\
\end{cases}
\end{equation}

\noindent
where $\psi=r/R_1$, $R_1=0.2$, $v_0=1$, $\rho_0=1, P_0=5$. We use $N=N^3_{1D}$ particles distributed in a glass-like configuration within a square box of length $L=1$. All simulations are carried out until $t=1$. The number of particles per axis $N_{1D}$ and the target number of neighbors $n_b$ are taken as input parameters. The quality of the simulations and convergence rate were checked with

\begin{equation}
    L_1=\frac{1}{N(r\le 0.5)}\sum_{i=1}^{N (r\le 0.5)}  \vert v_{\phi_i}^{a}-v_{\phi_i}^{s} \vert\,,
    \label{L1}
\end{equation}

\noindent where $v_{\phi}^a$ and $v_{\phi}^s$ stand for the analytical, Eq.~\ref{vortex_1}, and the simulated value of the azimuthal component of the velocity, respectively. Note that, unlike other works that calculate $L_1$ grouping the data first into bins to reduce the numerical noise \cite{spr10, hu2014, val16}, we directly use all SPH particles until distance $r=0.5$ in Eq.~\ref{L1}.

Figure~\ref{fig:Vortex_0} shows the last snapshots of the velocity distribution (top row) and its profile (bottom row) in the calculations with the highest resolution for the $C^6$ and $S^{0.9}_{4,9}$ interpolators. We see that both kernels give similar good results when the number of neighbors is high ($n_b=400$), but they notably differ when the number of neighbors is low ($n_b=100$) with a much larger dispersion of the azimuthal velocity in the $C^6$ calculation.

\begin{figure*}
\centering
\includegraphics[width=0.7\textwidth]{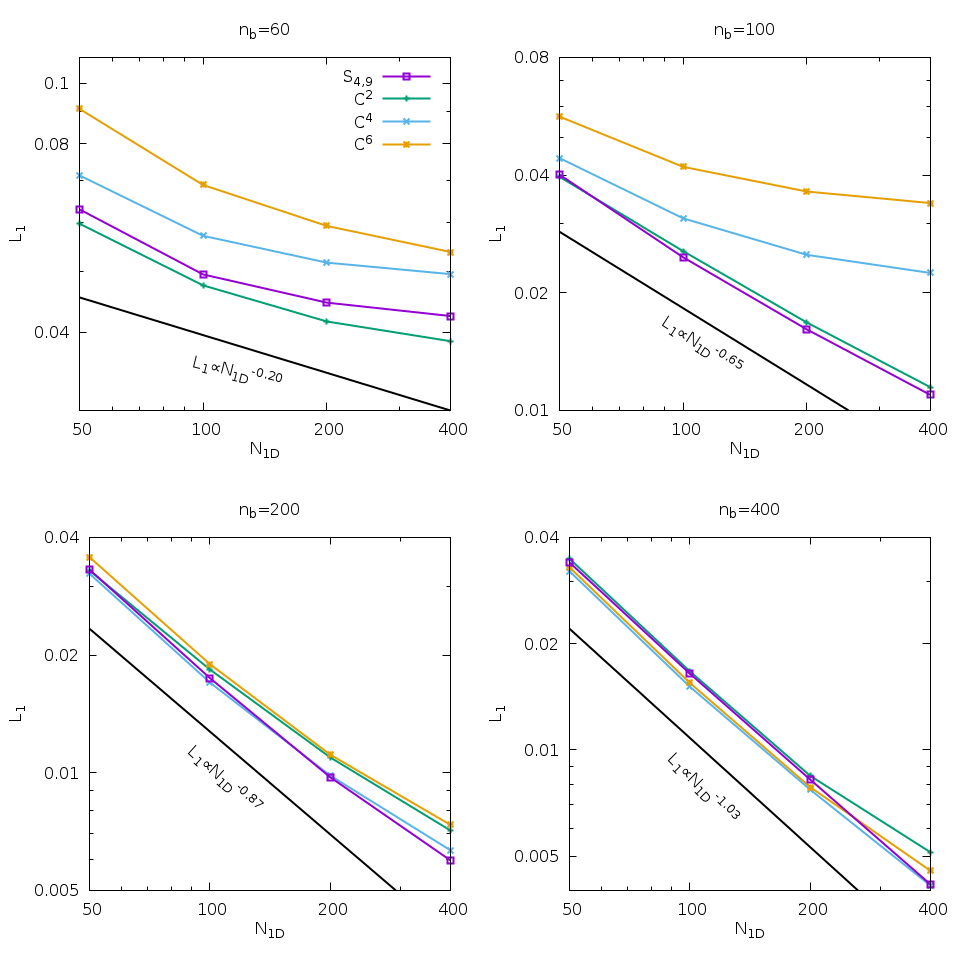}
\caption{Same as Figure \ref{fig:Vortex_1} but depicting the $L_1$ error as a function of $N_{1D}$ for different neighbor ($n_b$) choices. The solid black line shows $L_1\propto N_{1D}^{-p}$, where $p$ is the convergence rate and it has been obtained by adjusting a power law to the data of the $S^{0.9}_{4,9}$ curve.}
\label{fig:Vortex_2}
\end{figure*}

We show quantitative results on the behavior of the estimator $L_1$ in figures \ref{fig:Vortex_1} and \ref{fig:Vortex_2}. These cover a good portion of the $(N_{1D}, n_b)$ region commonly used in hydrodynamic calculations. Overall, our results match those of \cite{deh12} but the $L_1$ errors are lower, even when we use the $C^n$ polynomials. This reduction in $L_1$ is basically the result of the integral approach (IA) methodology, adopted in SPHYNX to calculate gradients, plus some other additional improvements \citep{garciasenz2022}. Figure~\ref{fig:Vortex_1} shows the profiles of $L_1$ obtained with different kernels as a function of the initial number of neighbors, $n_b$, at several $N_{1D}$, and at $t=1$. The dotted black line in the first two panels is for the pure $S_6$ which behaves similarly to $C^6$ for $n_b<150$. It is worth noting the rapid growth of the error when $n_b< 100$ for $C^2$ and $S_{4,9}$ and when $n_b\le 150-200$ for $C^4$ and $C^6$. The $L_1$ error is remarkably large for the interpolators $C^6$ and $S_6$ when $n_b\le 200$. These results agree with those of the previous test (Sect.~\ref{sec:first_test}) which were shown in the right column in Fig.~\ref{tensile_1}.

\begin{figure}
\centering
\includegraphics[width=\columnwidth]{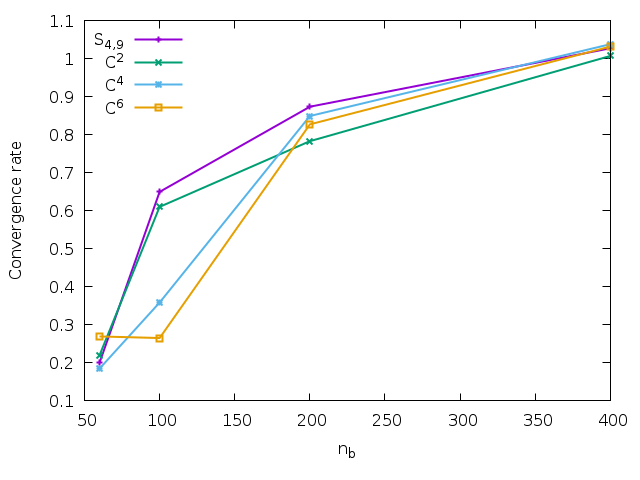}
\caption{Convergence rate (exponent $p$ in $L_1\propto N_{1D}^{-p}$) evolution for the Gresho-Chan test in function of the number of neighbors and for each kernel.}
\label{fig:convcomp}
\end{figure}
Figure~\ref{fig:Vortex_2} also shows the convergence to the analytical value of the azimuthal velocity at $t=1$ when both, the number of particles and neighbors rise. As stated in previous works \citep{zhu15} we find that good convergence is only attained for large enough combinations of $(N_{1D}, n_b)$. Nevertheless, the choice of the interpolator also matters, with $S_{4,9}^{0.9}$ providing the best results on average, in the entire range of $n_b$ considered in this work and for any $N_{1D}$. Among the Wendlands, the $C^2$ interpolator is the best when $n_b< 100$, but turns to $C^4$ in the interval $200\le n_b \le 400$. In general, the mixed Sinc gives the best results when $n_b\ge 100$, and still performs well when the number of neighbors is low, $n_b\simeq 60$, albeit slightly worse than $C_2$ for such low $n_b$. The absolute best result among all the calculated models, $L_1=0.0041$, was that of both $S_{4,9}^{0.9}$ and $C^4$ with $N^{1D}=400$, $n_b=400$ (bottom right panel), significantly lower than $L_1=0.0088$ previously reported by \citep{deh12} for a similar initial setting. It is similar to the best value, $L_1=0.0054$, obtained with the state-of-the-art relativistic SPH code by \cite{ros15} in 2D, with the $C^6$ kernel, and only when coupled with the integral approach to calculating the gradients. All panels in Fig.~\ref{fig:Vortex_2} depict the line $L_1\propto N_{1D}^{-p}$ (in black) with $p$ determined with the data of the mixed Sinc interpolator. The value of $p$ steadily increases with the number of target neighbors, from $p=0.20$ at $n_b=60$ to $p=1.03$ at $n_b=400$. This is clearly shown in Fig.~\ref{fig:convcomp}, where we present the convergence rate $p$ as a function of the number of neighbors for different kernels. Extrapolating, it suggests that the slope could be significantly higher than $p = 1$ provided that the number of neighbors is high enough, which is in agreement with \cite{zhu15}. On another note, the convergence rate with $p= 1.03$, shown in the last panel of Fig.~\ref{fig:Vortex_2}, is faster than that obtained in several previous calculations \citep{spr10,deh12} and similar to the value reported by \cite{ros15}. In general terms, Fig.~\ref{fig:convcomp} shows that the convergence rate of the mixed Sinc kernel is higher than that of the other represented kernels and certainly higher for the range $100\le n_b\le200$, commonly used in 3D SPH simulations.

\subsubsection{Fine-tuning the mix}
\label{subsubsec:tuning}
The results of the Gresho-Chan vortex test can be further tuned by changing the parameters in the mix, namely the Sinc exponents and the coefficient $\alpha$. Here we can benefit from the continuum nature of the Sinc kernels and choose among non-integer values of the exponents. Due to the many combinations, we restrict the exploration to $\alpha=0.90, 0.95$ which, according to Fig.~\ref{fig:zeroes}, show a reasonable balance between FWHM and $k_c$. A qualitative analysis can be performed in light of the adopted values of FWHM and $k_c$, which we assume as the main magnitudes driving the behavior of the interpolator.

\begin{figure}
\centering
\includegraphics[width=\columnwidth]{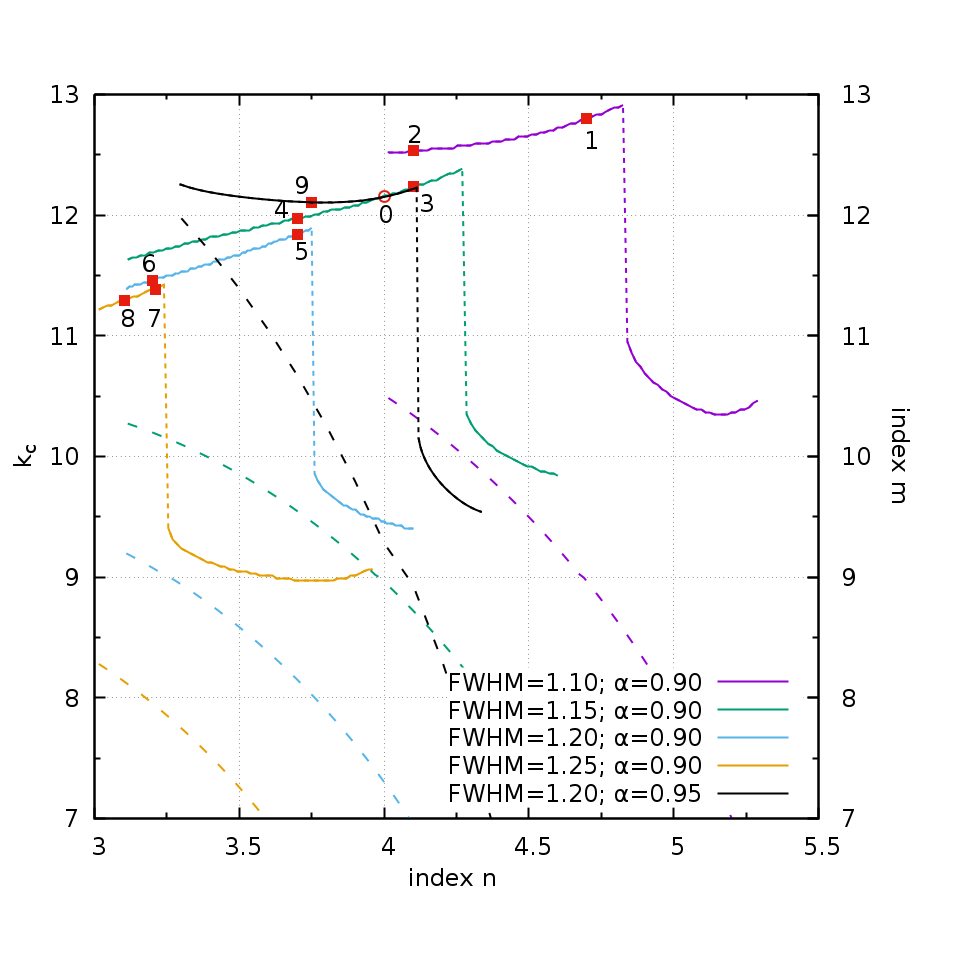}
\caption{Lines of constant FWHM of the Sinc kernels depicting the first zero of the Fourier Transform (continuum lines) and the Sinc exponents: $n$ (in the X-axis) and $m$ (dashed lines and the right Y-axis) in the mix with $\alpha=0.90$ (except the lines in black calculated with $\alpha=0.95$). The individual square points (in red) are those chosen in the fine-tuning study in Sect.~\ref{subsubsec:tuning} and Tables~\ref{table_fineT1} (points 0-8) and \ref{table_fineT2} (point 9). The circular point is for $S_{4,9}^{0.9}$, the kernel taken as default in this work and serves as a reference.}
\label{fig:FineTuning_1}
\end{figure}

First, we choose a value of FWHM around that of our reference kernel $S^{0.9}_{4,9}$ (FWHM=1.15). Next, we vary the first exponent of the Sinc within the interval $3\le n\le 5.5$ and then the second exponent, $m$, is obtained from the FWHM by solving the inverse problem. Once the pair $(n,m)$ is known, we can determine the first zero of its FT. The results are depicted in Fig.~{\ref{fig:FineTuning_1}} where each color corresponds to a particular value of the FWHM of the mixed kernel. We indicate the location of $S^{0.9}_{4,9}$ along the FWHM = 1.15 line with a red circle and the label 0, which serves as a reference. Next, we monitored the $L_1$ error of several particular points (red squares) on the lines with constant FWHM and compared them to the $L_1^0$ value obtained with $S^{0.9}_{4,9}$. The main results are shown in Table~\ref{table_fineT1}, where the anticorrelation of the FWHM and the Fourier critical wavenumber $k_c$ is evident. We note that the deviations around the reference value are lower than $15\%$. The combination with label 4 slightly enhances the results at low $n_b$ with minimal degradation of the others. The remaining cases follow the expected trend with better $L_1$ at large FWHM and low $n_b$ and vice versa. This suggests that previous knowledge of the number of neighbors may help to establish the optimal mix of kernel exponents. For example, combinations number 6 and 8 could be a good choice in the low-$n_b$ regime, but without losing too much of their pairing resistance in the high-$n_b$ regime.

On another note, raising the value of $\alpha$ slightly improves the results in the high$-n_b$ region. According to Fig.~\ref{fig:zeroes} the points on the $\alpha=0.95$ line (in yellow) located at $(4,9)$, $(4,10)$, $(4,11)$, and $(4,12)$ are good potential targets as they display FWHM values higher than those of $\alpha=0.9$ and very similar $k_c$ values to those of $\alpha=0.9$ (albeit slightly lower). The $L_1$ errors of these points are shown in Table \ref{table_fineT2} which also includes the $(3.75,10.46)$ combination, labeled as point 9 in Fig.~\ref{fig:FineTuning_1} and located just above points 4 and 5. For the most part, they do not improve over the default choice $S_{4,9}^{0.9}$ except when $n_b\simeq 400$, where there is a modest enhancement ($\lesssim5\%$). Figure~\ref{fig:FineTuning_1} also shows the line of constant $FWHM=1.20$ at $\alpha=0.95$ (black solid line). Interestingly, such a line intersects with $FWHM=1.15$ and $\alpha=0.90$ (green solid line) at the default choice (red circle). The $(3.75, 10.46)$ combination in Table~\ref{table_fineT2} (point 9 in Fig.~\ref{fig:FineTuning_1}) slightly reduces the normalized $L_1$ in all cases. It could be a plausible alternative to our default $S_{4,9}^{0.9}$.

On the same line, one may wonder if the mixing of two Wendland kernels of different orders would significantly enhance the results of the vortex test. To explore this issue, we carried out several additional simulations with the mix $C^{2,6}=0.9 C^2+ 0.1 C^6$ and compared the results with those obtained with pure Wendlands. The main results are shown in Table~\ref{table_fineT3} where we can see that, for the most part, the mixing only makes small changes in the $L_1$ values. A comparison between $L_1 (C^{2,6})$, $L_1(C^2)$, and $L_1(C^6)$ is shown in the last two columns of Table~\ref{table_fineT3}. The mixed Wendland slightly enhances the outcome of $C^2$ at low $n_b$ but the results in the high $n_b$ region are a bit worse. On another note, the $C^{2,6}$ kernel substantially improves results over the pure $C^6$ in the low-$n_b$ region but this fact is not a merit of the $C^{2,6}$ mix, but rather a demerit of the $C^6$ when $n_b< 200$. On the whole, the percent variations are similar to those shown in Tables~\ref{table_fineT1} and \ref{table_fineT2} with respect to the mix of Sinc kernels.

From the experiments above we conclude that the kernel $S_{4,9}^{0.9}$ is close to the optimal choice in the interval $60\le n_b\le 400$. Nevertheless, small to moderate improvements in particular $n_b$ regions are possible by carefully tuning $(n,m)$ and $\alpha$ as, for example, model 8 in Table \ref{table_fineT1} (low-$n_b$ enhancement) or any model in Table \ref{table_fineT2} (high-$n_b$ enhancement).

\begin{table*}
        \centering
        \caption{$L_1$ errors of selected points in Figure~\ref{fig:FineTuning_1} (square points) normalized to the $L_1^0$ errors obtained with the default $S^{0.9}_{4,9}$ (circular point) in the vortex test, with $\alpha=0.90$ and $N_{1D}=100$. The ratio is given for three representative numbers of neighbors, $n_b=60$, $200$, $400$ with $L_1^0=4.95~10^{-2}$, $1.75~10^{-2}$, $1.65~10^{-2}$, respectively.}
        \resizebox{16.5cm}{!}{
        \begin{tabular}{cccccccccc}
                \hline
                Point label & 0 & 1 & 2& 3&4&5&6&7&8\\
                (n,m)& (4.0,9.0) &(4.7,8.95)&(4.1,10.33)&(4.1,8.73)&(3.7,9.55)&(3.7,8.15)&(3.2,9.08)&(3.21,7.94)&(3.11,8.13)\\
                FWHM& 1.15&1.10&1.10&1.15&1.15&1.20&1.20&1.25&1.25\\
                $k_c/2\pi$& 1.935&2.037&1.996&1.946&1.905&1.884&1.824&1.813&1.797\\
                \hline
                \hline
                $(L_1/L_1^0)_{n_b=60}$ & 1.00& 1.14 &1.03 & 1.02&0.94&0.94&0.87&0.87&0.85\\
                $(L_1/L_1^0)_{n_b=200}$ & 1.00& 0.99 & 1.05& 0.98&1.02&0.99&1.03&1.03&1.04\\
                $(L_1/L_1^0)_{n_b=400}$ & 1.00& 0.97 & 1.00 & 0.99&1.01&1.02&1.04&1.10&1.09\\
                \hline
                \hline
        \end{tabular}}

        \label{table_fineT1}
\end{table*}

\begin{table*}
        \centering
        \caption{Normalized $L_1$ errors of several kernel combinations and $\alpha=0.95$ with respect to the default $S^{0.9}_{4,9}$ at $N_{1D}=100$ in the vortex test. The ratio is given for three representative numbers of neighbors, $n_b=60, 200, 400$.}
        \begin{tabular}{ccccccc}
                \hline
                (n,m)&(4.0,9.0)&(4.0,10.0)&(4.0,11.0)&(4.0,12.0)&(3.75,10.46) \\
                FWHM& 1.10&1.10&1.15&1.15&1.20 \\
                $k_c/2\pi$&1.932&1.946&1.989&2.421&1.926 \\
                \hline
                \hline
                $({L_1}/{L_1^0})_{n_b=60}$ &1.01  &1.02 & 1.02&1.03& 0.99 \\
                $(L_1/L_1^0)_{n_b=200}$ &0.94  &0.96& 0.98& 1.01& 0.98\\
                $(L_1/L_1^0)_{n_b=400}$ & 0.95 & 0.96&0.96&0.96&0.97\\
                \hline
                \hline
        \end{tabular}

        \label{table_fineT2}
\end{table*}

\begin{table*}
        \centering
        \caption{Comparison of the $L_1$ errors of pure $C^2$, $C^6$, and the combination $C^{2,6}=\alpha C^2+(1-\alpha) C^6$ with $\alpha=0.9$ of Wendland kernels in the vortex test.}
        \begin{tabular}{cccccccc}
                \hline
                \multirow{2}{*}{$N_{1D}$}&  \multirow{2}{*}{$n_b$}& $L_1(C^2)$& $L_1(C^6)$&$L_1(C^{2,6})$& \multirow{2}{*}{\Large{$\frac{L_1(C^{2,6})}{L_1(C^2)}$}}&  \multirow{2}{*}{\Large{$\frac{L_1(C^{2,6})}{L_1(C^6)}$}}\\
                & &\scriptsize{$(\times 10^{-2})$}&\scriptsize{$(\times 10^{-2})$}&\scriptsize{$(\times 10^{-2})$}& &\\
                \hline
                \hline
                $50$ & $60$ & 5.96&9.10 &5.62& 0.94 &0.62\\
                $50$ & $200$  &3.29 & 3.56& 3.33& 1.01&0.94\\
                $50$ & $400$ &3.48 &3.29 &3.51 & 1.01&1.07\\
                $100$ & $60$ &4.75  & 6.88&4.45 & 0.94 &0.65\\
                $100$ & $200$  &1.84 & 1.89 &1.80  & 0.98&0.95\\
                $100$ & $400$ &1.67 & 1.54&1.64 &1.02&1.06\\
                \hline
                \hline
        \end{tabular}

        \label{table_fineT3}
\end{table*}

\subsection{Shock-tube test}
\label{sec:third_test}
The numerical experiment known as the shock-tube test involves the propagation of shock and rarefaction waves in an inhomogeneous medium. In this test, a box of size $(1\times0.125\times0.125)$ is filled with a gas so that there is an initial strong contrast in density and pressure between the left $(x<0.5)$ and right ($x> 0.5)$ regions of the box. This test has an analytical solution \citep{sod1978} to compare with. Hence, we can perform the $L_1$ analysis of the results in light of the adopted kernel and is therefore adequate for the present study. Unlike precedent tests, the sharpness of the kernel function is relevant here if we want to resolve the shock front and the contact discontinuity. This slightly favors high-order kernels.

The onset is similar to that adopted in other works \citep{Wadsley17, price18} with $\rho_l(x<0.5)=1$, $p_l(x<0.5)=1$, and $\rho_r(x>0.5)=0.125$, $p_r(x>0.5)=0.10$ in the left and right states of the box, respectively. Both regions of the 3D box are filled with particles of equal mass settled in a glass-like configuration with $N_l=8.64\cdot 10^5$ and $N_r=1.08\cdot 10^5$ particles on the left and right sides of the box, respectively. The initial density profile around the contact discontinuity was not smoothed. The large initial density and pressure jumps drive a strong shock, which makes the artificial viscosity (AV) algorithm especially relevant in this test. Previously published works \citep{Wadsley17} obtained good results in this test, free or almost free of velocity oscillations in the post-shock plateau, by choosing relatively high AV parameters: $\alpha_{max}=2, \beta=4$, in combination with AV switches \citep{cul10} or with lower values, but constant, $\alpha=1$ and $\beta=2$ \citep{price18}. In this work, we decided to use the approach by \cite{Wadsley17} with the same choice of AV parameters.

\begin{figure*}
\centering
\includegraphics[width=\textwidth]{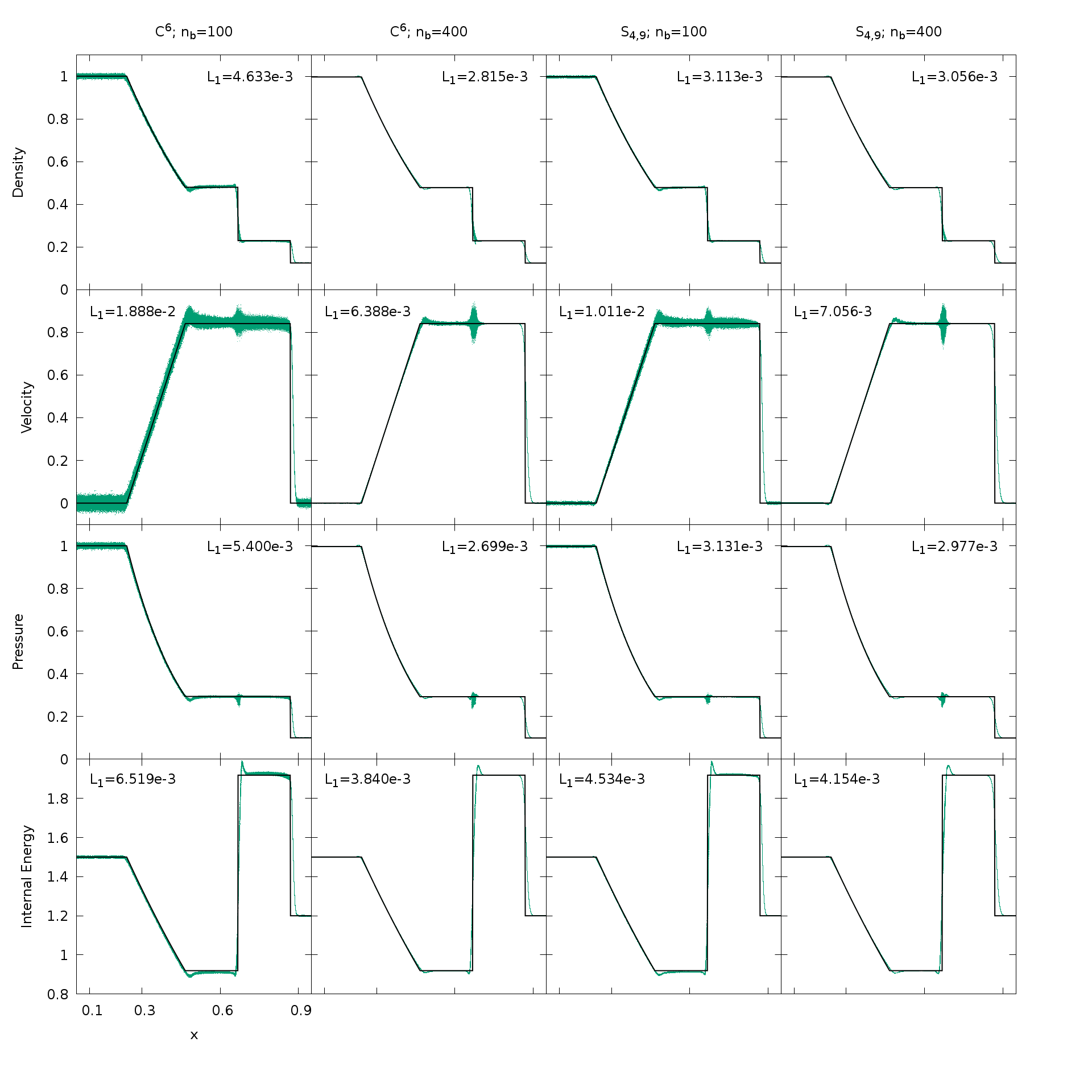}
\caption{From top to bottom, profiles of density, shock velocity, pressure, and internal energy in the 3D shock-tube test. From left to right, the columns show the results for the $C^6$ and $S_{4,9}^{0.9}$ kernels for two different neighbor counts each. All SPH particles are plotted. Each panel reports the corresponding $L_1$ values with respect to the theoretical solution, shown in solid black lines.}
\label{fig:Sodprofiles}
\end{figure*}

Figure~\ref{fig:Sodprofiles} shows the results of this test for kernels $C^6$ and $S_{4,9}^{0.9}$, with neighbor counts of $n_b=100, 400$. In general, at large neighbor count $C^6$ exhibits slightly better results than $S_{4,9}^{0.9}$, as expected for a higher order kernel, but the differences are not large. Furthermore, $S_{4,9}^{0.9}$ significantly outperforms $C^6$ at a lower neighbor count, which can be seen in the lower particle dispersion and lower $L_1$ values. This is consistent with the results found in previous tests, where $S_{4,9}^{0.9}$ shows a robust performance across a broad spectrum of neighbor counts.

\begin{figure*}
\centering
\includegraphics[width=0.7\textwidth]{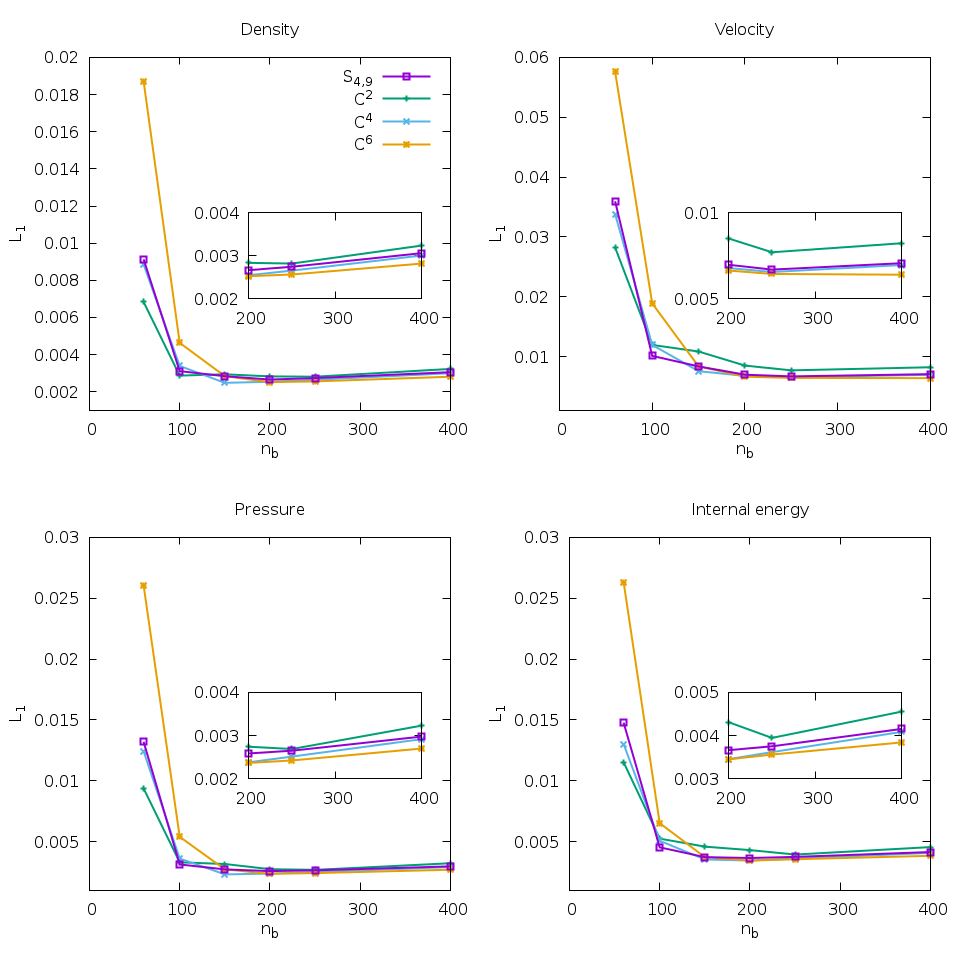}
\caption{$L_1$ error evolution in function of the number of neighbors for different kernels and magnitudes. Each panel includes an inset zooming in the region of large neighbor count.}
\label{fig:Sod}
\end{figure*}

The same behavior can be seen in Fig.~\ref{fig:Sod}, which shows how the $L_1$ error evolves with increasing number of neighbors, for the three Wendland kernels and $S_{4,9}^{0.9}$, in the four physically relevant magnitudes. In general, both $C^4$ and $S_{4,9}^{0.9}$ provide low errors for a wide range of neighbor counts. On the other hand, the benefits of $C^6$ can only be seen when using large numbers of neighbors, and the improvement in $L_1$ is only marginal. The opposite is true for the Wendland $C^2$ which becomes competitive only at low neighbor counts, $n_b\le 100$.

\section{Conclusion}

We reviewed the properties of the Sinc family of interpolators proposed by \cite{cabezon2008} and extended them to consider the impact of having a binary linear mix of these kernels. Low-order Sincs have good sampling properties, but they are prone to suffering from pairing instability when the number of neighbors is high. In contrast, high-order Sincs are resistant to pairing, but suffer from bias when the number of neighbors is low\footnote{Which is also true for other interpolators, such as the B-splines and Wendland kernels.}. We show that an appropriate choice of the mixing coefficient considerably delays the critical wavenumber $k_c$ at which the Fourier transform of the kernel becomes negative, making them much more resistant to pairing instability while increasing the FWHM at the same time, which makes them less prone to bias. In this respect, the combination of low- and high-exponent Sinc functions not only produces new, spherically symmetric, and pair-resistant interpolators but also improves their internal sampling. Having good sampling properties, even when the number of neighbor particles is low or moderate, is a very desirable quality for any kernel.

The particular combination of $n=4$ and $n=9$ Sincs with mixing coefficient $\alpha=0.9$, namely $S_{4,9}^{0.9}=0.9 S_4+ 0.1 S_9$, proved to be very suitable as an interpolator for SPH simulations. For example, it accurately reproduces the density of a homogeneous system in {\it both}, ordered lattices and glass-like particle distributions in a wide range of neighbor particles (Sect.~\ref{sec:zeroth_test}). According to the results in Sect.~\ref{sec:first_test}, it also hinders the spawning of the pairing instability when the number of neighbors is large, $n_b\simeq 400$. Additionally, it maintains a good partition of the unit and an accurate $\left<\Delta r\right> \simeq 0$ condition even when the number of neighbors is low, $n_b\simeq 60$. The analysis of the results of the Gresho-Chan test in Sect.~\ref{sec:second_test} allows us to make a detailed study of the convergence issues of the SPH scheme. This test is appropriate to contrast the quality of the results in light of the chosen kernel, either the Sinc $S_{4,9}^{0.9}$ or the widely used $C^2$, $C^4$, and $C^6$ Wendland kernels. We found that the behavior of $S_{4,9}^{0.9}$ was better on average than that of commonly used Wendland polynomials in the explored range $50\le N_{1D}\le 400$, each with $60\le n_b\le 400$. The only exception is when the number of neighbors is very low, $n_b\simeq 60$ for which the $C^2$ polynomial led to the lowest $L_1$ values. Our results also suggest that working with $n_b\ge 200$, or $\eta \ge 1.8$ in Eq.~\ref{eta}, would make the simulations less dependent on the nature of the chosen kernel and provide quite accurate results. All our results start to flatten around that value (see Figs.~\ref{fig:static_box}, \ref{tensile_1}, \ref{fig:Vortex_1}, and \ref{fig:convcomp}). Still, interpolators that can give good results even at $n_b\simeq 100$, such as $S_{4.9}^{0.9}$ and $C^2$, are valuable as they increase local resolution and reduce computational burden.

We also tested the performance of $S_{4.9}^{0.9}$ in the presence of strong shocks simulating the Sod tube in 3D. In this case, higher-order, peaked kernels are expected to be favored for properly resolving the discontinuities. Indeed, $C^6$ produces the lowest errors, but only for large neighbor counts and at the expense of losing spatial resolution. Wider, less peaked kernels offer a better trade-off for low and medium numbers of neighbors, making $C^4$ and $S_{4.9}^{0.9}$ a better choice. These results confirmed the conclusions drawn in the Gresho-Chan test, establishing $S_{4.9}^{0.9}$ as a versatile kernel, with good interpolation properties in a wide range of numbers of neighbors, and in different physical scenarios.

Summarizing, we showed that a suitable linear mix of Sinc functions, combining low and high exponents, produces new, spherically symmetric interpolators with both, better internal sampling and enhanced pair-resistant behavior. The mixed Sincs are well balanced, allowing them to work with very different initial numbers of neighbors, which can, for example, be applied in different fluid regions with the same kernel. Moreover, it makes the SPH method more robust, because the results become less sensitive to statistical fluctuations in the number of interpolating neighbor particles. Although this strategy can be used to enhance any other kernel as, for example, the popular $M_n$ and $C^n$ polynomials, it is well suited to the Sincs because of their continuum nature, which endows them with great flexibility. According to Fig.\ref{fig:zeroes} it is enough to carefully choose the exponents $n,m$ ($n<m$) and the mixing coefficient $\alpha$ so that particle clustering is avoided once the initial number of neighbors has been chosen. Incorporating such a mixture of kernels into an SPH code is straightforward and encompasses little overall difference in computational burden. This is because the mixed interpolator and its gradient can be tabulated and loaded at the beginning of the calculation. It makes therefore feasible to work with non-integer Sinc exponents to fine-tune the results, as shown in Sect. \ref{subsubsec:tuning}.

Although the results of this work point to a very positive impact of combining kernels of a different order, work has to be done to extend the analysis to different physical situations than those considered in this work, for example, instability growth and turbulence simulation. A robust interpolator, capable of coping with both, low and high numbers of neighbor particles, may help to balance the resolution among fluid regions with very different densities without the need of having particles with different masses. This is an interesting line of development that will make the SPH technique even more adaptive.

\section*{Acknowledgements}

This work has been supported by the MINECO Spanish project PID2020-117252GB-100 grant SGR-386/2021, by the AGAUR/Generalitat de Catalunya (DG) and by the Swiss Platform for Advanced Scientific Computing (PASC) project SPH-EXA2: Smoothed Particle Hydrodynamics at Exascale (RC). The authors acknowledge the support of the center for scientific computing: \mbox{sciCORE} (\url{http://scicore.unibas.ch}) at the University of Basel, where part of these calculations was performed.

\section*{Data Availability}

Data supporting this article will be shared on reasonable request with the corresponding author. A current version of SPHYNX is publicly available at:~\url{https://astro.physik.unibas.ch/sphynx}


\bibliographystyle{mnras}
\bibliography{bibliography_r}


\bsp	
\label{lastpage}
\end{document}